\documentclass[9pt,sigconf]{acmart}

\usepackage{tikz}
\usepackage[most]{tcolorbox}
\usepackage{gensymb}
\usepackage{todonotes}
\usepackage{amsmath,amsfonts}
\usepackage{algorithm}
\usepackage[noend]{algpseudocode}
\usepackage[symbol]{footmisc}

\usepackage{tikz}
\usepackage{amsmath}
\usepackage{multirow}
\usepackage{booktabs}

\newcommand*\circled[1]{\tikz[baseline=(char.base)]{
            \node[shape=circle,fill,inner sep=1pt] (char) {\textcolor{white}{#1}};}}
\usepackage{xcolor}
\usepackage{pgfplots}
\usepackage{tikz}
\pgfplotsset{compat=1.8}
\pgfplotsset{
    width=\textwidth,
}
\usepackage{lipsum}
\usepackage{pgfplotstable}
\usetikzlibrary{pgfplots.groupplots}

\usepackage{tikz}
\usepackage{hyperref}
\hypersetup{
    colorlinks=true,
    linkcolor=blue,
    filecolor=magenta,      
    urlcolor=cyan,
    pdftitle={Overleaf Example},
    pdfpagemode=FullScreen,
    }

\definecolor{codegreen}{rgb}{0,0.6,0}
\definecolor{codegray}{rgb}{0.5,0.5,0.5}
\definecolor{codepurple}{rgb}{0.58,0,0.82}
\definecolor{mGreen}{rgb}{0,0.6,0}
\definecolor{mGray}{rgb}{0.5,0.5,0.5}
\definecolor{mPurple}{rgb}{0.58,0,0.82}
\definecolor{backcolour}{rgb}{0.95,0.95,0.92}

\definecolor{RYB1}{RGB}{80, 99, 42}
\definecolor{RYB2}{RGB}{215, 227, 191}
\definecolor{RYB3}{RGB}{198, 187, 174}
\definecolor{RYB4}{RGB}{146, 205, 220}
\definecolor{RYB5}{RGB}{238, 144, 34}
\definecolor{RYB6}{RGB}{142, 172, 59}

\pgfplotscreateplotcyclelist{colorbrewer-RYB}{
{RYB1!50!black,fill=RYB1},
{RYB2!50!black,fill=RYB2},
{RYB3!50!black,fill=RYB3},
{RYB4!50!black,fill=RYB4},
{RYB5!50!black,fill=RYB5},
{RYB6!50!black,fill=RYB6},
}

\usepackage{lscape} 
\usepackage{pdflscape}
\usepackage{afterpage}
\usepackage{capt-of}
\usepackage{listings}
\usepackage{float}
\usepackage{array,multirow,graphicx}
\usepackage[titletoc]{appendix}
\usepackage{caption}
\usepackage{subcaption}
\usepackage{soul}
\usepackage{pifont}
\usepackage{svg}

\definecolor{ggreen}{HTML}{2CC225}
\definecolor{yyellow}{HTML}{C2C80A}
\definecolor{bbrown}{HTML}{8e4603}

\definecolor{codegreen}{rgb}{0,0.6,0}
\definecolor{codegray}{rgb}{0.5,0.5,0.5}
\definecolor{codepurple}{rgb}{0.58,0,0.82}
\definecolor{mGreen}{rgb}{0,0.6,0}
\definecolor{mGray}{rgb}{0.5,0.5,0.5}
\definecolor{mPurple}{rgb}{0.58,0,0.82}
\definecolor{backcolour}{rgb}{0.95,0.95,0.92}

\lstdefinestyle{CStyle}{
    commentstyle=\color{mGreen},
    keywordstyle=\color{magenta},
    numberstyle=\tiny\color{mGray},
    stringstyle=\color{mPurple},
    basicstyle=\sffamily\footnotesize,
    frame=lrtb,
    breakatwhitespace=false,         
    breaklines=true,                 
    captionpos=b,                    
    keepspaces=true,                 
    numbers=left,                    
    numbersep=5pt,                  
    showspaces=false,                
    showstringspaces=false,
    showtabs=false,                  
    tabsize=2,
    language=C
}

\lstdefinestyle{CStyle1}{
    commentstyle=\color{mGreen},
    keywordstyle=\color{magenta},
    numberstyle=\tiny\color{mGray},
    stringstyle=\color{mPurple},
    basicstyle=\sffamily\footnotesize,    frame=lrtb,
    breakatwhitespace=false,         
    breaklines=true,                 
    captionpos=b,                    
    keepspaces=true,                 
    numbers=left,                    
    numbersep=5pt,                  
    showspaces=false,                
    showstringspaces=false,
    showtabs=false,                  
    tabsize=2,
    language=C
}

\lstdefinestyle{mystyle}{
    commentstyle=\color{codegreen},
    keywordstyle=\color{magenta},
    numberstyle=\tiny\color{codegray},
    stringstyle=\color{codepurple},
    basicstyle=\sffamily\footnotesize,
    breakatwhitespace=false,         
    breaklines=true,                 
    captionpos=b,                    
    keepspaces=true,                 
    numbers=left,                    
    numbersep=5pt,                  
    showspaces=false,                
    showstringspaces=false,
    showtabs=false,                  
    tabsize=2,
    language=C
}
\lstdefinestyle{trans}{
    commentstyle=\color{codegray},
    numberstyle=\tiny\color{codegray},
    stringstyle=\color{codepurple},
     basicstyle=\sffamily\footnotesize,
    frame=lrtb,
    breakatwhitespace=false,         
    breaklines=true,                 
    captionpos=b,                    
    keepspaces=true,                 
    numbers=left,                    
    numbersep=5pt,                  
    showspaces=false,                
    showstringspaces=false,
    showtabs=false,                  
    tabsize=2,
     language=[x86masm]Assembler,  escapeinside={\%*}{*)},   
     }     

\AtBeginDocument{%
  \providecommand\BibTeX{{%
    \normalfont B\kern-0.5em{\scshape i\kern-0.25em b}\kern-0.8em\TeX}}}

\copyrightyear{2024}
\acmYear{2024}
\setcopyright{acmlicensed}\acmConference[ICCAD '24]{IEEE/ACM International
Conference on Computer-Aided Design}{October 27--31, 2024}{New York, NY,
USA}
\acmBooktitle{IEEE/ACM International Conference on Computer-Aided Design
(ICCAD '24), October 27--31, 2024, New York, NY, USA}
\acmDOI{10.1145/3676536.3676822}
\acmISBN{979-8-4007-1077-3/24/10}

\begin{document}

\title{LaserEscape: Detecting and Mitigating Optical Probing Attacks }

\author{Saleh Khalaj Monfared}
\affiliation{%
  \institution{Worcester Polytechnic Institute}
       \city{Worcester}
  \state{MA}
    \country{USA}
}
\email{skmonfared@wpi.edu}

\author{Kyle Mitard}
\affiliation{%
  \institution{Worcester Polytechnic Institute}
         \city{Worcester}
  \state{MA}
    \country{USA}
}

\email{krmitard@wpi.edu}

\author{Andrew Cannon}
\affiliation{%
  \institution{University of Florida}
         \city{Gainesville}
        \state{FL}
      \country{USA}
 }
\email{andrew.cannon@ufl.edu}

\author{Domenic Forte}
\affiliation{%
  \institution{University of Florida}
         \city{Gainesville}
        \state{FL}
      \country{USA}
 }
\email{dforte@ece.ufl.edu}

\author{Shahin Tajik}
\affiliation{%
  \institution{Worcester Polytechnic Institute}
       \city{Worcester}
  \state{MA}
    \country{USA}
  }
\email{stajik@wpi.edu}

\renewcommand{\shortauthors}{Khalaj Monfared et al.}

\begin{abstract}

The security of integrated circuits (ICs) can be broken by sophisticated physical attacks relying on failure analysis methods. Optical probing is one of the most prominent examples of such attacks, which can be accomplished in a matter of days, even with limited knowledge of the IC under attack. Unfortunately, few countermeasures are proposed in the literature, and none have been fabricated and tested in practice. These countermeasures usually require changing the standard cell libraries and, thus, are incompatible with digital and programmable platforms, such as field programmable gate arrays (FPGAs). In this work, we shift our attention from preventing the attack to detecting and responding to it. We introduce \textit{LaserEscape}, the first fully digital and FPGA-compatible countermeasure to detect and mitigate optical probing attacks. \textit{LaserEscape} incorporates digital delay-based sensors to reliably detect the physical alteration of the fabric caused by laser beam irradiations in real time. Furthermore, as a response to the attack, \textit{LaserEscape} deploys real-time hiding approaches using randomized hardware reconfigurability. It realizes 1) moving target defense (MTD) to physically move the sensitive circuity under attack out of the probing field of focus to protect secret keys and 2) polymorphism to logically obfuscate the functionality of the targeted circuit to counter function extraction and reverse engineering attempts. We demonstrate the effectiveness and resiliency of our approach by performing optical probing attacks on protected and unprotected designs on a 28-nm FPGA. Our results show that optical probing attacks can be reliably detected and mitigated without interrupting the chip's operation.

\end{abstract}





\settopmatter{printfolios=true}

 \maketitle

 \section{Introduction}\label{sec:introduction}
In various applications, the security of a computing machine relies on sensitive assets stored on Root-of-Trust integrated circuits (ICs).
Examples of these on-chip assets include cryptographic keys, proprietary firmware, and intellectual property (IP).
Attackers can extract these assets by gaining physical access to the devices where they are stored and mounting physical attacks.
Attacks relying on failure analysis (FA) tools pose the biggest threat against ICs.
Prominent examples of such attacks are the various forms of optical probing attacks~\cite{lohrke2016no,tajik2017power,chef2018descrambling,krachenfels2021real,krachenfels2021automatic} which enable the adversary to recover secret data and functions from chips.
Illuminating a laser beam on transistors and measuring their states using the modulated backscattered laser beam is the main mechanism behind these optical probing attacks. 

Deploying randomization in countermeasures, such as masking~\cite{nikova2006threshold,gross2016domain}, is a conventional method to mitigate conventional side-channel attacks, as it prevents the repetition and integration of the measurements over multiple clock cycles.
However, randomness becomes ineffective if the adversary performs optical probing on halted circuits or data at rest~\cite{krachenfels2021real,krachenfels2021automatic}.
Moreover, many root/master keys, which cannot be masked, are vulnerable to optical probing during key caching from memory to registers~\cite{lohrke2016no,krachenfels2021automatic}.

A few prevention and detection-based countermeasures have been proposed in the literature to avert optical probing attacks.
These countermeasures try to protect the design either directly from laser irradiation~\cite{rahman2021concealing,parvin2023hidden,zhang2023laser} or detect tampering with the voltage/clock, which is required for certain optical probing attacks~\cite{farheen2022twofold}.
Such schemes deploy various logic styles and design techniques to either minimize the data/function-dependent leakage through the backscattered laser light~\cite{rahman2021concealing,parvin2023hidden} or to detect the generated heat illumination of the laser beam on logic cells~\cite{zhang2023laser}.
Most of these solutions require the modification of standard cell libraries and are incompatible with legacy systems and field programmable gate arrays (FPGAs).
Moreover, they are tailored toward a specific form of optical probing and, thus, cannot mitigate the threat in general.
Finally, these schemes have yet to be fabricated or tested in practice.

The only FPGA-compatible sensing countermeasure that has been proposed so far deploys ring-oscillators (ROs)~\cite{tajik2017pufmon} to detect laser irradiation.
However, such a solution has never been researched further due its high sensitivity to environmental conditions, which leads to high false alarms. 
In this work, we attempt to answer the following research questions: \emph{(1) Is it possible to confidently detect optical probing attempts in real-time using fully digital components of an FPGA? (2) What would be a suitable response upon detecting an optical probing attempt to deceive the adversary and not interrupt the IC's operation?}


\noindent\textbf{Our Contribution.} To answer the above questions, we present \textit{LaserEscape}, which is capable of detecting and responding to optical probing attacks using the existing resources of mainstream FPGAs.
To detect an optical probing attempt, we repurpose the recently proposed fully digital delay-based 1LUTSensor~\cite{Jayasinghe_Udugama_Parameswaran_2023}, which had been originally been designed for remote power analysis side-channel attacks.
We found that the mechanism which allows 1LUTSensor to detect power fluctuations with only one look-up table (LUT) can also detect laser irradiation from the IC backside silicon.
Afterward, to respond to the attack, we explore novel moving target defense and gate polymorphism strategies to deceive the adversary in real-time and render the attack ineffective.
First, we demonstrate that how upon the detection of the attack, the partial reconfiguration (PR) feature of mainstream FPGAs and programmable SoCs can be utilized to dynamically randomize the placement of security-critical registers and moving them out from attacker's probing field with a low overhead to protect secret keys.
Second, we show how polymorphic logic gates can be realized on LUTs of FPGAs to zeroize the output of the logic function to prevent function extraction by the attacker.
Finally, we apply \textit{LaserEscape} to various circuits on a 28-nm FPGA and perform optical probing on them to assess the effectiveness of our solution.

\noindent\textbf{Source Code Availability.} We publicly publish our raw sensor data and make the source code of \textit{LaserEscape} open source in \url{https://github.com/vernamlab/LaserEscape}

 \section{Technical Background}\label{sec:Background}

\subsection{Optical Probing Attacks}\label{opticalSCA}

\begin{figure*}[!t] 
\centering
\includegraphics[width=0.8\linewidth]{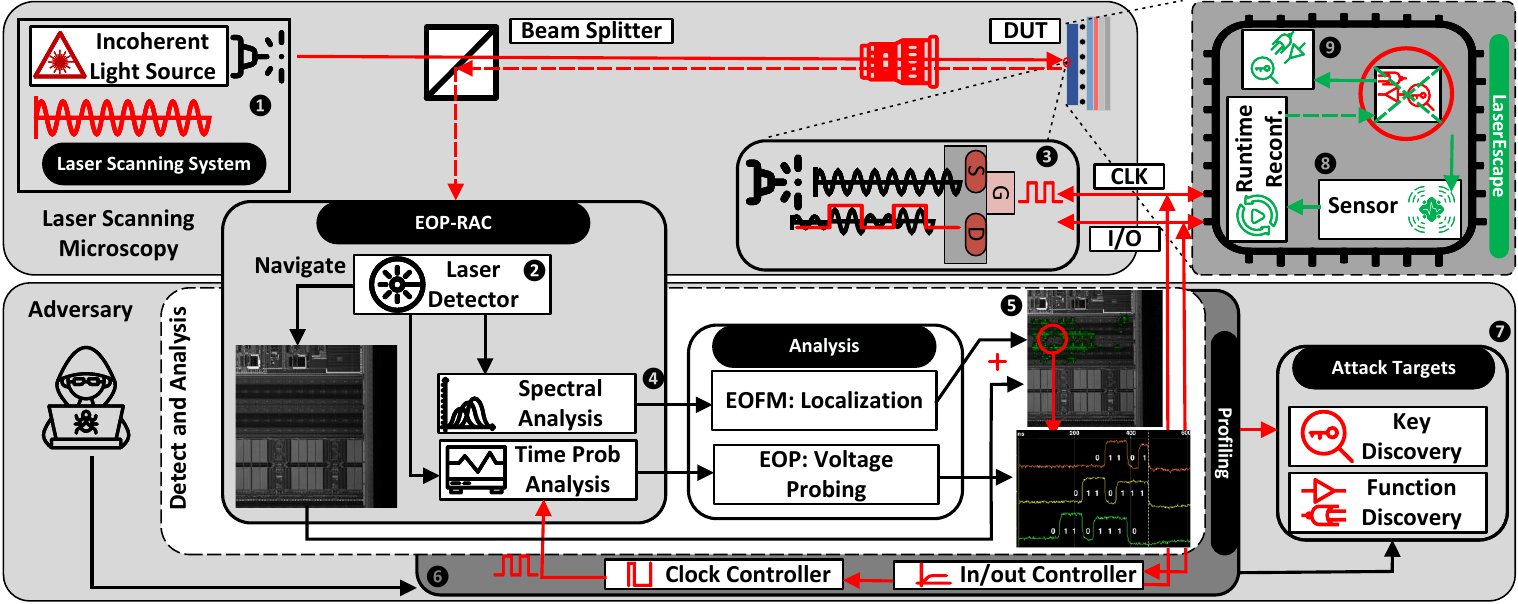}
\caption{High-level overview of Optical Attacks and \textit{LaserEscape} mitigation}
\label{overview}
\end{figure*}

Fig.~\ref{overview} illustrates a high-level overview of optical probing attacks.
As illustrated, a laser scanning microscope can be used to navigate through a target IC as well as capture dynamic transistor-level activities on it. 
Optical probing attacks often include extracting sensitive data from the target IC while it is operational~\cite{lohrke2016no,tajik2017power,chef2018descrambling,krachenfels2021automatic,krachenfels2021real}.
For this aim, a high-power incoherent light source (HIL) with a wavelength of 1.3 $\mu$m is utilized. 
In \circled{1}, the light beams are illuminated and passed through multiple optical mirrors, including scanning mirrors, before entering the objective lenses. 
The illumined beams are reflected from the IC, and are guided to the laser detector in \circled{2} using splitters. 
By focusing the objective lens onto the region of interest (RoI), a 2D image of the target area can be captured and then be used for localizing physical components on the IC. Additionally by moving the stage different parts of the IC can be navigated.

Once focused, HIL beams can effectively pass through the silicon layer and, as shown in \circled{3}, reflect. 
The run-time alterations in the transistor-level components modulate reflected beams. 
In \circled{4}, these beams are captured, amplified, and prepared in the analog domain. 
As the received signals contain the activities of the scanned area, they can be used to localize components that operate at a certain target frequency using electro-optical frequency mapping (EOFM)(\circled{5}). 
Furthermore, by focusing on a particular node on the IC, high precision electro-optical probing (EOP)\footnote{When using a coherent light source, EOP is typically called Laser Voltage Probing~(LVP), and EOFM is called Laser Voltage Imaging~(LVI).} can be mounted to estimate gate-level voltage fluctuations in pico-second time accuracy. 
At this point, the adversary can interpose the pattern image and activity signals to physically localize the targets on the IC as shown in \circled{5}.
It is common that the adversary can perform multiple profiling attacks at this stage (\circled{6}) by controlling I/O, global clock, and reset of the IC, and gain information about the target.
For instance, by resetting the IC with a certain frequency or by injecting a particular pattern of inputs into the IC~\cite{tajik2017power}, the target elements could be identified using EOFM.

\subsection{Sensor Implementations on FPGAs}\label{fluc}
In the case of laser attacks, such as optical probing, the colliding laser photons locally increase the temperature of the target wires and transistors on the chip.
These local temperature variations can affect the propagation delays of electrical signals in delay-based sensors on FPGAs, such as ring-oscillators and time-to-digital converters (TDCs).
ROs~\cite{zhao2018fpga} and TDCs~\cite{schellenberg2021inside} have been extensively used for remote side-channel attacks on FPGAs, where they measure voltage fluctuations on the power delivery network (PDN) caused by adjacent intellectual property cores.
While these sensors are often used for offensive side-channels, they can also be used for the detection of active attacks causing temperature and current variations.
RO networks have been employed as the first attempt to detect optical probing attacks on FPGAs~\cite{tajik2017pufmon}.
The main drawbacks of using RO networks are their high power consumption and sensitivity to environmental conditions, leading to a high number of false alarms\cite{tajik2017pufmon}.
Therefore, as described later, a more reliable TDC candidate will be chosen to detect optical probing attacks. %

\subsection{Placement Randomization as MTD}\label{pr}
To perform optical probing, the placement of the circuit under attack has to be constant for the duration of the attack, as the laser beam needs to stay for a minimum amount of time on each transistor to reveal the state of the transistor.
Hence, if the placement of a circuit can be dynamically changed or randomized, optical probing becomes infeasible due to the reduced signal-to-noise ratio (SNR).
To achieve randomized layouts for a circuit, the dynamic reconfiguration feature of mainstream FPGAs, such as partial reconfiguration (PR) or redundant logic and multiplexers on application specific ICs (ASICs) can be used.
The PR feature of SRAM-based FPGAs represents a unique capability that allows for the dynamic alteration of specific sections of the FPGA while maintaining the uninterrupted operation of the remaining system. This functionality not only boosts the system's flexibility but also optimizes power usage and enhances adaptability to evolving operational conditions or demands~\cite{koch2012partial1}. Additionally, PR's application in FPGAs has proven valuable in mission-critical settings where system downtime is impermissible and real-time processing is necessary~\cite{koch2012partial}. 


AMD/Xilinx’s dynamic function exchange (DFX) introduces a technology for setting up PR areas within a static system, allowing users to allocate modules to these specific areas on FPGA fabrics~\cite{xilixreconfig}.
However, the Vivado toolchain has several weaknesses, such as being too slow for real-time applications and the lack of support for bitstream relocation~\cite{manev2022byteman}.
On the other hand, the open-source tools, such as Byteman~\cite{manev2022byteman}, improved the efficiency and speed of performing PR from an embedded processor, making it suitable for real-time applications, such as countermeasures against side-channel attacks during runtime.

Several countermeasures against side-channel attacks, like power and electromagnetic (EM) analysis, have already leveraged PR~\cite{mentens2008power,guneysu2011generic,heyszl2012localized,moradi2013comprehensive,hettwer2019securing,bow2020side,khan2021moving}.
Some of these approaches use PR to introduce timing jitter, which can help distort power traces and defeat power side-channel attacks. 
Other techniques involve dynamically relocating functional units to different regions of the FPGA through PR, dissociating the spatial correlation between EM emanations and the target logic.
The capability for real-time reconfiguration of FPGAs through PR has also been explored to create side-channel resistant implementations for cryptographic primitives, such as~\cite{monfared2024randohm}.

\subsection{Gate-Level Polymorphism as MTD}\label{polymorphic}
Another approach to respond to an optical probing attempt, in the case of function extraction, is to transform the gate functionality.
Such real-time function transformation can be obtained using polymorphic circuits.
Polymorphic circuits are digital circuits with the ability to perform two or more logical operations, where the selected operation depends on external factors such as temperature, light, or supply voltage. 
Polymorphic circuits were first introduced in CMOS circuits~\cite{stoica2001polymorphic} using evolutionary algorithms. 
The classic example of a polymorphic circuit is a gate whose output functionality is OR when operating at high supply voltage, but switches to AND at low supply voltage. 
Further works have explored static design methodologies for polymorphic gates without the need for evolutionary algorithms \cite{bernard2021design}. 
These circuits can be used in CMOS circuits to change pipeline functionality with low area impact and are compatible with a wide variety of logic function combinations.

While most of the polymorphic gate realizations are carried out by custom cell libraries, multiplexers and switch boxes have also been used to design polymorphic elements~\cite{Falkinburg_2011}.
In this work, we leverage the flexibility of LUTs within FPGAs for the implementation of polymorphic circuits. 
LUTs can be programmed with the union of the truth tables of two different logic functions, with one input to the LUT selecting which logical function is propagated to the output. 
By combining this LUT-based approach with external sensors, gate-level polymorphism can be achieved without the need for reprogramming LUTs or rerouting the circuit.

\section{Threat Model and Attack Scenarios}

For optical probing attacks, we presume that an adversary is equipped with FA tools and aims to extract a secret key or reverse engineer specific elements. 
The primary motivation for these attacks comes from the scenario where a single key is employed across all devices, such as in instances where firmware, bitstream, or logic encryption secures the proprietary design of a system. 
The vendor programs this key into the product before distribution to customers. Thus, acquiring the key from one device could compromise the security of the entire device family. Moreover, even if individual devices use different keys, the uniform layout across a device family means that once an adversary masters key extraction from one device, she can apply this knowledge to breach other devices within the same family.
At the right side of the Fig.~\ref{overview}, attack scenarios are highlighted in \circled{7}. For this article we consider optical-based 1) key recovery attacks\cite{krachenfels2021automatic,krachenfels2021real}, 
and 2) function recovery attempts\cite{mehta20241}.

To evaluate each of these scenarios, we consider suitable optical probing techniques. Specifically, we consider both EOP and EOFM for key recovery, and for function reversing, we consider an EOFM attack.
Then, we deploy \textit{LaserEscape} prototype on the target device and verify our mitigation methodology accordingly. 
As displayed in Fig.~\ref{overview}, our proposed detection and mitigation methodologies are highlighted with \circled{8} and \circled{9}. 
We implement a low-overhead delay-based sensor in proximity to the target elements. 
We assume that the sensor is supplied by an internal clock source so that the adversary cannot tamper with it.
Upon the illumination of the HIL at the protected physical zone, the physical characteristics of the target are temporary altered and sensed by the proposed detector in \circled{8}. 
This triggers the defense response on the DUT to modify the target element. 
In \circled{9} the indicated elements are modified, removed, or redeployed based on the strategy of the defender in real-time.

 \section{Attack Detection}


\subsection{Sensor Requirements}
As mentioned in Sect.~\ref{fluc} optical probing attacks induce temperature fluctuations within the chip. 
These localized variations alter the signal propagation delays of the transistors. 
An ideal sensor must satisfy multiple conditions to detect such attacks:

\noindent\textbf{Large Spatial Coverage.} Performing optical probing over a specific area only increases the temperature locally in the wires and transistors of that region. 
Therefore, an ideal sensor should provide coverage across the entire security-sensitive areas to detect localized temperature variations.

\noindent\textbf{Large Temporal Coverage.} The irradiation of individual gates or registers during an optical probing attempt is potentially an extremely rapid process (e.g., less than 1 ms). 
Hence, the sensor must have a higher time resolution and continuity to capture an attack successfully.


\noindent\textbf{Reliability.} The sensor must be immune to voltage and temperature variations, so to detect the attack under various environmental conditions reliably.
Otherwise, the sensor could detect noise as an attack attempt and generate false alarms.

\subsection{Sensor Architecture}\label{sec:sensor}
\begin{figure}[!t]
    \centering
    \includegraphics[width=0.40\textwidth]{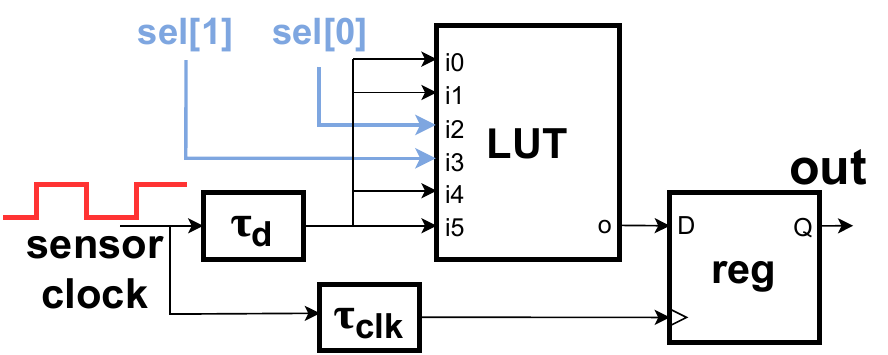}    \caption{Block diagram of 1LUTSensor}
    \label{fig:1lut_blockdiagram}
\end{figure}

Based on the above requirements, we will present the architecture of our delay-based sensors in this section.
Our sensor design is inspired by the 1LUTSensor originally implemented for on-chip power analysis attacks implemented in~\cite{Jayasinghe_Udugama_Parameswaran_2023}, see Fig.~\ref{fig:1lut_blockdiagram}. 
It consists of a clock signal connected to the data and clock inputs of a register simultaneously.
Moreover, the delay of the clock and data paths can be tuned using PVT (Process, Voltage, Temperature) invariant delay elements, namely one or multiple IDELAYE2 elements found in the I/O blocks of the AMD 7 series FPGAs (or similar tapped delay elements in other FPGA families).
The signal on the register's data input path passes through a LUT, in which the path from the LUT inputs to the output can be configured for further adjustments.

\begin{figure}[!t]
  \begin{subfigure}[b]{0.155\textwidth}
    \includegraphics[width=\linewidth]{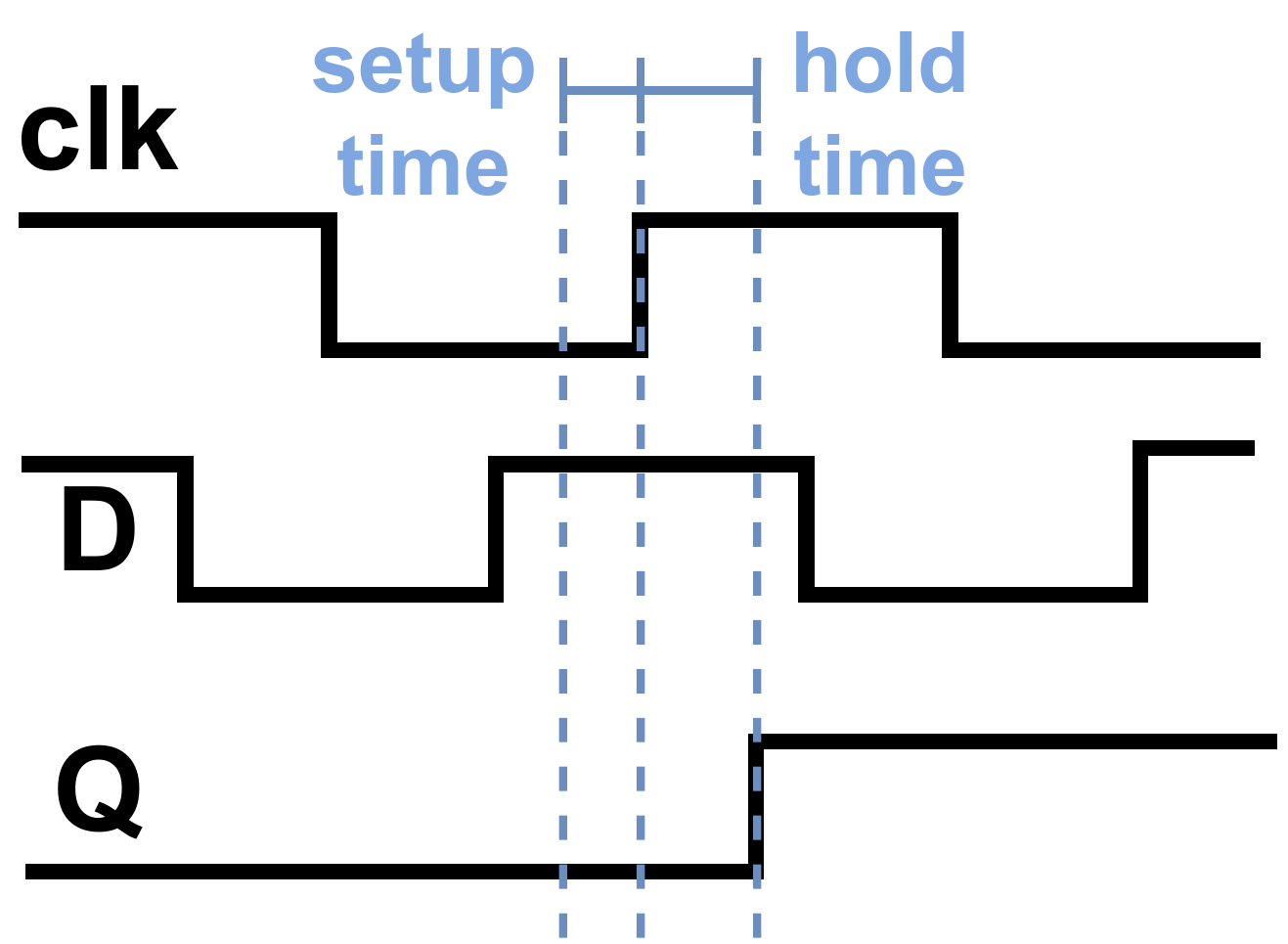}
         \caption{Constant 1 case}
         \label{fig:1lut_timing_1}
     \end{subfigure}
     \hfill
  \begin{subfigure}[b]{0.155\textwidth}
    \includegraphics[width=\linewidth]{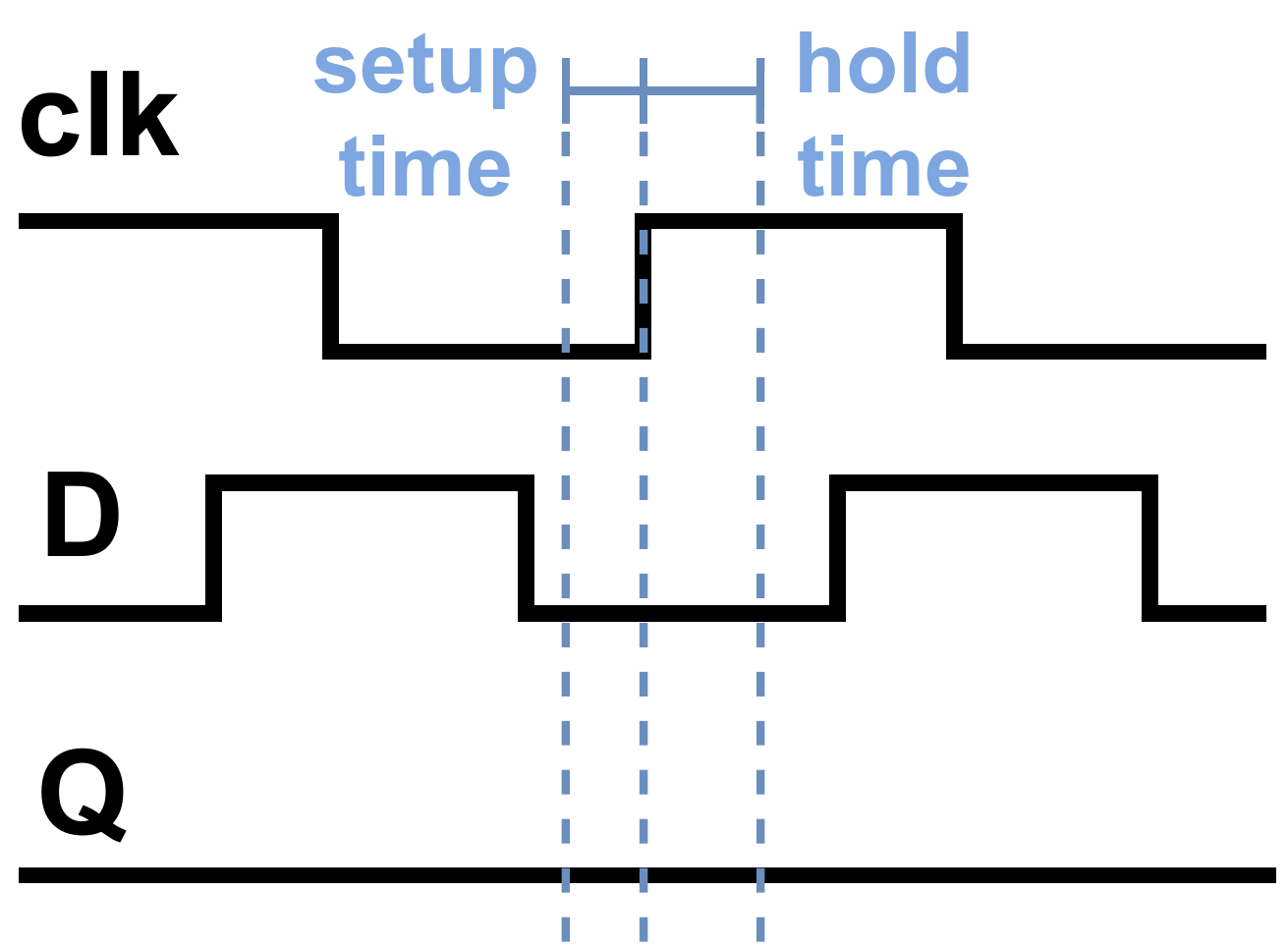}
         \caption{Constant 0 case}
         \label{fig:1lut_timing_0}
     \end{subfigure}
     \hfill
  \begin{subfigure}[b]{0.158\textwidth}
    \includegraphics[width=\linewidth]{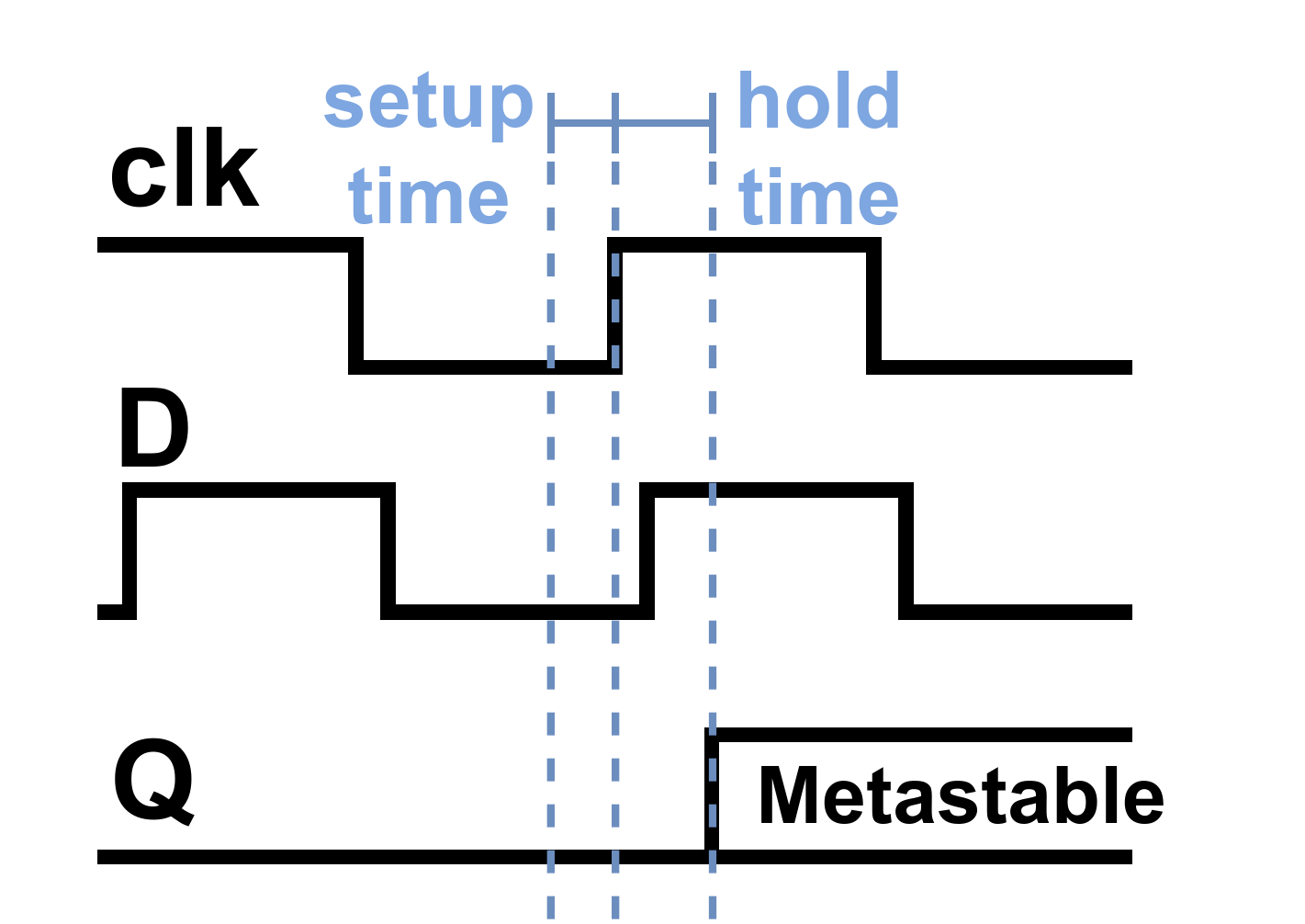}
         \caption{Metastable case}
         \label{fig:1lut_timing_m}
     \end{subfigure}
    \caption{Timing diagrams of 1LUTSensor}
    \label{fig:1lut_timing}
\end{figure}

At runtime, we set the delays of each path and select the input of the LUT to pass through such that the data pulse arrives a short moment after the setup time boundary, as shown in Fig.~\ref{fig:1lut_timing}. 
In this case, in principle, the sensor output stays at 1. 
By increasing the delay of the LUT, the output moves toward metastablity. 
We can validate this behavior by iterative sampling the output of the register and observing more 0 occurrences.  
As a result, by finding the right tuning levels, the sensor could be stabilized by outputting 1 and getting triggered (outputting 0) when its delay is increased due to an increase in temperature.

The primary reason behind choosing this sensor is its size compared to other TDCs, in that it only requires one LUT and can be placed in any slice, satisfying the spatial coverage requirements.
Furthermore, the utilization of these PVT-invariant elements for delay tuning makes the sensor immune to environmental conditions, satisfying its reliability.
Finally, supplying the sensor a few tens of megahertz clock signal provides sufficient timing resolution less than the laser dwell time on a pixel.\\

A key difference in our implementation is using a much lower clock frequency. The original 1LUTSensor as described in \cite{Jayasinghe_Udugama_Parameswaran_2023} was developed to measure voltage fluctuations in a circuit running on the chip to perform power analysis attacks. As such, it ran as fast as it could on the FPGA fabric - 600MHz - to measure an AES circuit running at frequencies ranging from 12-120MHz with enough resolution to detect the effects of switching gates. We do not want such a precise resolution, since this would cause the circuit we wish to protect to add noise to our readings. In addition, the laser pulses for LVI are on the order of milliseconds, which means the fluctuations in the sensor may be of that size.

\subsection{Adjustable Delay Elements}

Tapped delay elements are extensively utilized in FPGAs to facilitate signal delays. These elements are capable of adjusting the delay of a wire in run-time. Different delay blocks are available for different FPGA families. For instance, AMD provides IODELAY1, IODELAY2, and IODELAY3 blocks in Spartan 6, 7 Series and UltraScale/+ families, respectively, featuring different adjustable delay blocks.~\cite{Jayasinghe_Udugama_Parameswaran_2023} 
Here, we describe the details of our sensor based on IDELAYE2 elements. However, the technique presented here can be extended to other delay blocks as well.

\begin{figure}[!t]
    \centering
    \includegraphics[width=0.45\textwidth]{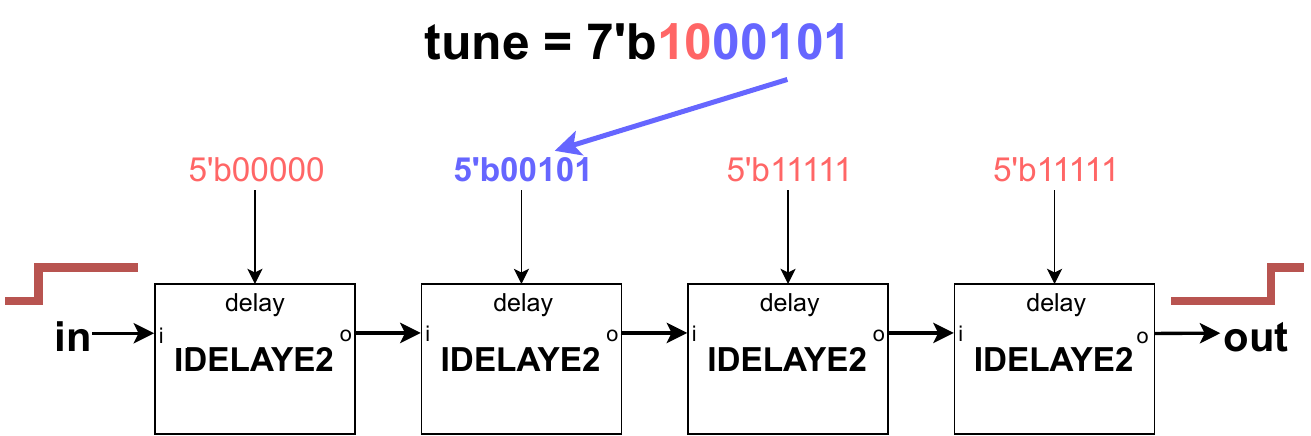}
    \caption{Block diagram of a 4-long IDELAYE2 chain}
    \label{fig:idelaye2_chain_blockdiagram}
\end{figure}

IDELAYE2 elements are 31-tap, wraparound, delay elements with a calibrated tap resolution~\cite{xilinx_idelaye2_2023}. 
Each IDELAYE2 acts as a pass-through for a 1-bit signal with a 5-bit input for the delay configuration, where the higher value leads to a larger delay.
Since the delay of the clock path is larger than that of the data path, a single IDELAYE2 might not provide sufficient delay.
Therefore, a chain of IDELAYE2s for the clock path can be used as shown in Fig.~\ref{fig:idelaye2_chain_blockdiagram}.

\begin{figure}[t]
    \centering
    \includegraphics[width=0.49\textwidth]{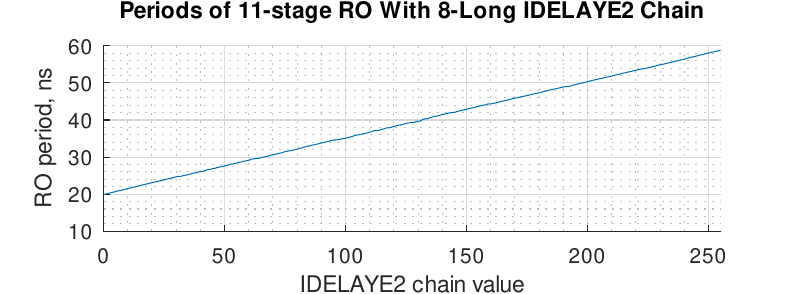}
        \caption{Graphs of ring oscillator period vs delay in IDELAYE2}
        \label{fig:ro_idelay_graphs}
\end{figure}

In our design, the chain consists of $2^n$ IDELAYE2s, where $n$ is any positive integer, and the delay input is $n+5$ bits wide.
The five least significant bits set the delay of one of the IDELAYE2s, and the $n$ most significant bits set how many of the rest have the maximum delay (11111). 
The rest have the minimum delay (00000). 
For example, in the 4-long chain shown in Fig.~\ref{fig:idelaye2_chain_blockdiagram}, $n=2$, so the tuning input is 7 bits wide. 
Similarly, if the delay were 1010111, one IDELAYE2 would have a delay of 10111, two IDELAYs would have the maximum delay, and one would have the minimum delay.
Moreover, we can verify the behavior of these delay blocks by including them in series with a RO on an AMD Kintex 7 FPGA. In this case, we can keep track of the RO's frequency based on the IDELAYE2 delay(s). Furthermore, by plotting the corresponding chain period as a function of delay values, we can determine the range and resolution of these delay blocks for fine-tuning the sensibility. 
Fig.~\ref{fig:ro_idelay_graphs} plots the period of an exemplary 11-stage RO chain with  8 IDELAYE2 blocks, showing $ps$ resolution for the sensor.\\

\subsection{Sensor Tuning}
In our sensor implementation, we use a standalone IDELAYE2 for the data path delay and an $m$-long IDELAYE2 chain for the clock path. 
Moreover, we use a select input to determine which input of the LUT goes through to the register.
Hence, we establish a \textit{tune value}, consisting of three values: 1) data-path delay, 2) clock-path delay, and 3) LUT select.
For many combinations of data path delay and LUT select, there is a corresponding value for the clock path delay where the output is metastable.
However, some combinations perform better since their accumulative zero counts are significantly smaller than others, which means they are metastable in a way that is closer to the constant zero case shown in Fig.~\ref{fig:1lut_timing_1}.
For every value of the data path delay, the binary search of clock delay for metastability is carried out with LUT select fixed at zero.
If metastability is found, we search for adjacent clock delay values with every LUT select value for other metastable tunes.
During this process, we keep track of the best tune, determined by the lowest maximum zero count over the $t_{sense}$ interval. At the end of the procedure, we extract the best tune.

\subsection{Sensor Output Post Processing} \label{sec:sensorpostprocessing}
When the sensor's output is metastable, yielding an output of $1$ with occasional zero pulses, the length and frequency of these pulses increase during optical probing due to the impact of increased temperature on delay lines. Hence, we utilize measurement circuitry to characterize this. First, we implement a \textit{zero counter}, which counts the number of clock cycles for which the sensor value is zero over $t_{detect}$. Secondly, we employ a \textit{pulse counter}, which counts the length of the zero pulses.
Based on the readings of the counters, we connect the sensor output to a latch, which we use to trigger our chosen defenses: 1) partial reconfiguration or 2) polymorphic gates, which will be described in Sect~\ref{sec:framework_movingtargetdefense} and Sect~\ref{sec:framework_gatepolymorphism}, respectively.









 \section{Attack Response}\label{sec:vprobe}

Real-time response against physical attacks can be addressed with different approaches in accordance to the threat model. 
For instance one can deploy a self-destructive mechanism to zeroize the sensitive data upon sensing the physical attacks~\cite{cannon2023protection,tada2021design}.
In contrast to solutions which necessitates cell library modification at pre-silicon~\cite{shen2018nanopyramid}, we propose fully logic-level mitigation methodologies that provides non-interrupting circumvention in addition to existing CMOS friendly zeroization~\cite{tada2021design,cannon2023protection}. 
Specifically, in our method target circuitry can benefit from real-time reconfiguration to bypass the optical attack and almost uninterrupted, operate as normal. 
This is useful for deceiving the adversary~\cite{tajik2017pufmon}. 
Furthermore, the full zeroization of the device including the entire memory contents incurs a huge area and latency for large designs,~\cite{cannon2023protection,srivastava2019efficient} which can be replaced by partial zeroization.
In addition to MTD, we incorporate a gate polymorphic design, particularly to defeat  optical reverse engineering attempts. We design the polymorphic gates by inspiring the state-of-the-art methodology in~\cite{cannon2023protection} to create a supply voltage-controlled gates and obfuscate the functionality of the target elements in real-time.


\subsection{Moving Target Defense via PR}\label{sec:framework_movingtargetdefense}
\begin{figure}[!t]
   \centering \noindent
   \includegraphics[width=0.8\linewidth]{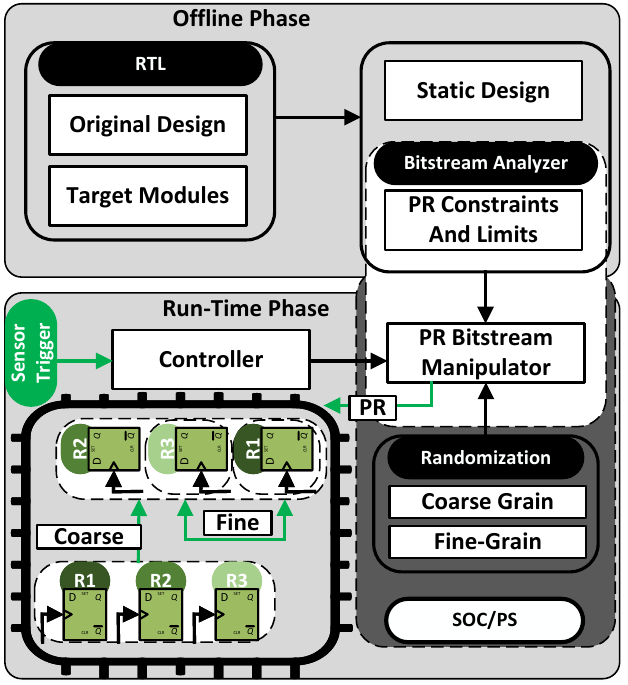}
	\caption{The Operation flow of the proposed Moving Target Defense in an FPGA.}
   	\label{fig:pr}
 \end{figure} 

Fig.~\ref{fig:pr} illustrates the design of our proposed MTD-based response for an optical attack on an FPGA. 
We consider an original hardware design which includes some sensitive data requiring protection. 
For simplicity, we assume the key registers $R_1$, $R_2$, and $R_3$ as the secret elements in our design. 
Using TCL, the targeted modules are specified by the user through predefined annotations in register-transfer level (RTL). 
At this stage, the static design in the original bitstream as well as sample partial files are generated.
The constraints such as the allowable range for FPGA slices, regional locations, and possible range of flip-flops (FFs) \cite{consXilinx} are established for generating reconfiguration.
This information, together with the generated bitstreams, are moved to the online phase of the framework. 
In this phase, a lightweight processing unit (PS) is employed within the SoC (or in an external processor) to execute reconfiguration. 
A secure, one-time pseudo random number generator (PRNG)~\cite{tsoi2003compact,monfared2020bsrng} is implemented to provide randomness to the PR-generator unit each time. 
The PR-generator includes a bitstream manipulator (Byteman\cite{manev2022byteman} in our instance), uses the predefined constraint limits to randomized location (LOC) and shuffle the chosen registers. These two randomization introduce coarse and fine-grain randomness to the reconfiguration, respectively. 
Consequently, the corresponding partial reconfiguration bitstream is created based on the original bitstream. 
The FPGA is then reprogrammed as soon as the trigger signal from the sensor is received.


\noindent\textbf{Inter-Register Hiding.}
The core functionality of the proposed online phase is bitstream manipulation aimed at generating one-time PR. 
This approach introduces real-time randomization, deviating from previous methods where bitstreams were stored in memory~\cite{khan2021moving}. This technique not only enhances the security level of the countermeasure but also reduces the memory usage of the PR files to a single bitstream. 
As depicted in Fig.~\ref{fig:pr}, the PRNG unit on the processor generates one-time randomness, and the bitstream manipulator program (e.g., Byteman) collects this randomized constraint information to produce one-time bitstream. 
In the coarse grain randomization, one or multiple slices in the FPGA are considers. All the corresponding elements in the target slices (e.g., target FFs and LUTs) are identified in the bitstream. 
The bitstream manipulator activates the corresponding routing and elements on another slices in a randomized possible reconfigurable module on the fly and disables the original slices of the target module and the FPGA proceeds to reconfiguration. 
The reconfiguration process delay overhead is considerably less than EOFM/EOP integration time, leading to successful mitigation and deceiving the attacker. These timing are investigated in detail in Sect.~\ref{sec:eval}. 

\noindent\textbf{Intra-Register Hiding.}
For fine-grain manipulation, we utilize DFX to implement PR as a hardware-based scrambling method to counter Laser Voltage Probing attacks. For this purpose, a high-level script (such as Python) is used to randomly permute target registers (e.g., the 3 bits of a master key shown in Fig.~\ref{fig:pr}). The potential search space for such permutations is super-exponential ($n!$), which effectively defeat probing attacks. This sequence is then transferred to the bitstream manipulator program to be used as constraints within the bitstream codes. Unlike the previous method, this approach minimizes resource utilization by requiring only a single reconfigurable module  for implementation. This efficiency arises because the only circuitry modification in this method involves the hardware referencing of the target flip-flops, which effectively alters the local internal routing in the target slice.

 \begin{figure}[!t]
     \centering
     \begin{subfigure}{0.18\textwidth}
        \centering
        \includegraphics[width=\textwidth]{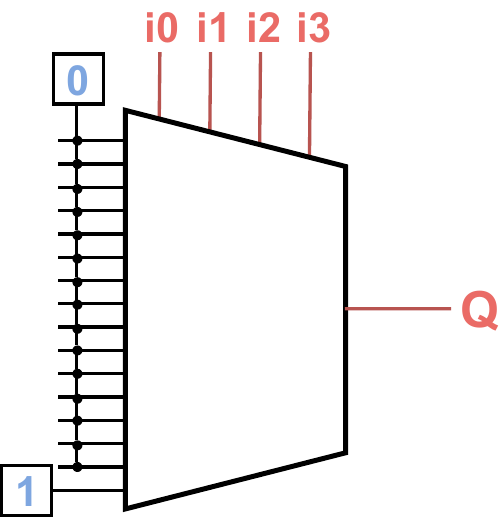}
        \caption{As one 4-input LUT}
        \label{fig:6in_lutx1_blockdiagram}
     \end{subfigure}
     \hspace{15pt}
     \begin{subfigure}{0.17\textwidth}
        \centering
        \includegraphics[width=\textwidth]{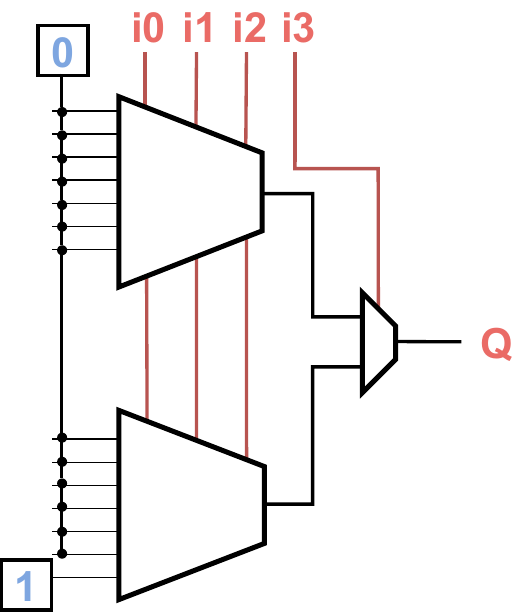}
        \caption{As two 3-input LUTs}
        \label{fig:5in_lutx2_blockdiagram}
     \end{subfigure}
        \caption{Block diagram of a 4-input LUT implementing a 4-input AND gate}
        \label{fig:various_lut6_blockdiagrams}
\end{figure}

\subsection{Gate Polymorphism} \label{sec:framework_gatepolymorphism}
To implement a polymorphic gate, we exploit the LUTs on the FPGA.
An $n$-input LUT consists of $2^n$ SRAM cells feeding into a $2^n$-to-1 multiplexer. One could formulate the truth table of an $n$-input logic function to the SRAM cells and use the select inputs as the inputs of the logic function, choosing which SRAM cell to pass on to the output. 
An example of a 4-input LUT implementing an AND gate is in Fig.~\ref{fig:6in_lutx1_blockdiagram}. Alternatively, this 4-input LUT can be seen as two 3-input LUTs with a 2-to-1 multiplexer to select which input is passed through to the output, as shown in Fig.~\ref{fig:5in_lutx2_blockdiagram}.

In general, for $n$-input LUTs, we can implement polymorphic $n-1$-input LUTs using a control input to change the logic function. For instance, we can exploit input $i3$ in Fig.~\ref{fig:5in_lutx2_blockdiagram} to switch the output between 1) The result of a 3-input AND and 2) constant zero. This idea can be generalized to implement other logic gates (i.e., XOR) easily by selecting proposer LUT inputs. Furthermore, by connecting the control input pin to the latch output of the sensor described in Sect.~\ref{sec:sensor}, we can modify any protected logic upon detecting optical probing attacks.

\section{Experimental Setup}

\begin{figure}[!t]
  \begin{subfigure}[b]{0.37\columnwidth}
    \includegraphics[width=\linewidth]{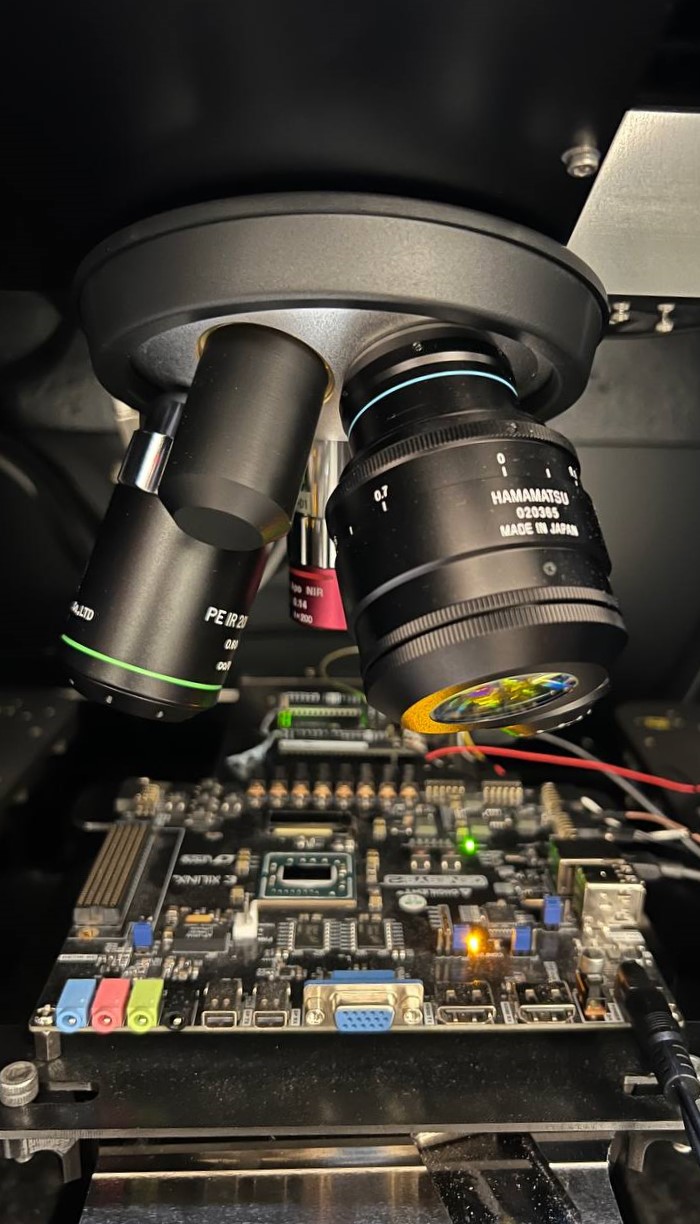}
    \caption{}
    \label{fig:experiment_set}
  \end{subfigure}
  \hspace{5pt} 
  \begin{subfigure}[b]{0.51\columnwidth}
    \includegraphics[width=\linewidth]{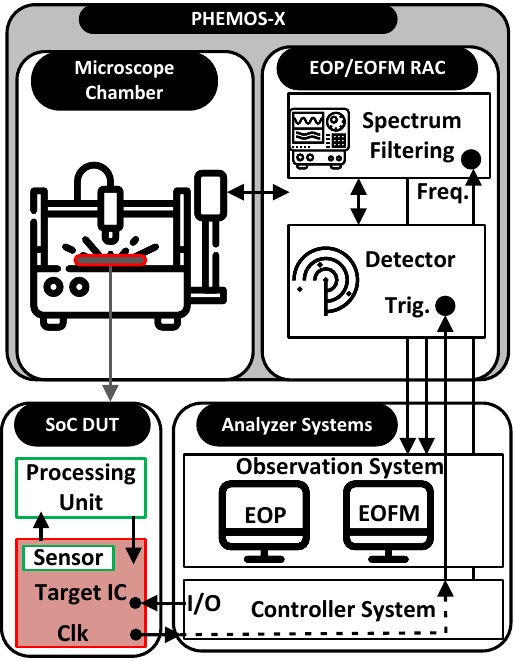}
    \caption{}
   \label{setup_zoomout}
  \end{subfigure}
    \caption{Measurement setup. (a) Objective lenses focusing on the target IC (b) High-level setup diagram.}
  \label{fig:experiment_set1}
\end{figure}



\textbf{Electrical and Optical Equipment.}\label{subsec:Measrement_setup}
A Hamamatsu PHEMOS-X FA microscope was used to perform optical probing, see Fig.~\ref{fig:experiment_set1}.
It is equipped with a 1.3 $\mu$m high-power laser source used for our analysis. Additionally objective lenses of 5X/0.14NA, 20X/0.6NA, 50X/0.76NA, and 71X are employed in our experiments. 
Furthermore, optical zooms of 2x, 4x and 8x were higher magnifications. 
As illustrated in Fig.~\ref{fig:experiment_set}, in the EOFM process, the DUT is scanned by a laser through the use of galvanometric mirrors. 



  \begin{figure*}[!t] 
     \centering
     \begin{subfigure}{0.95\textwidth}
        \centering
        \includegraphics[width=\textwidth]{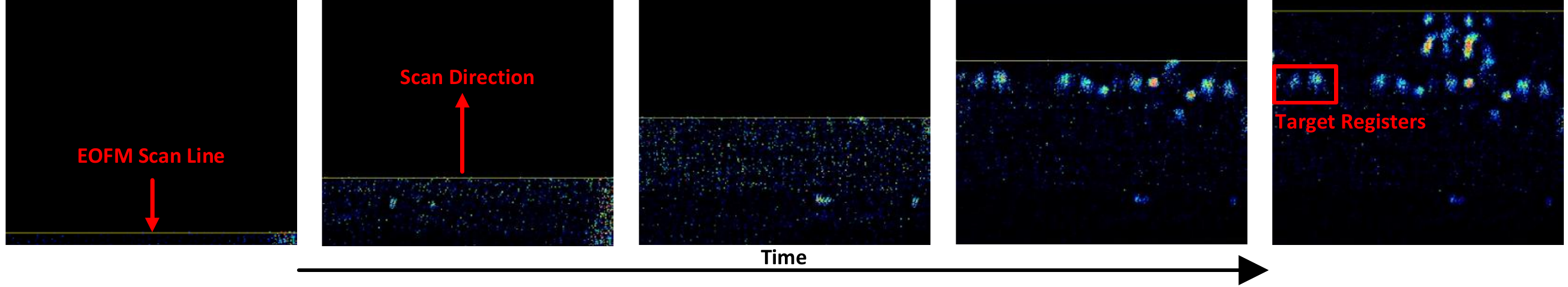}
        \vspace*{-6mm}
        \caption{timeline illustration of EOFM scan to extract targeting register values}
        \label{fig:eofm_un}
     \end{subfigure}
     \hspace{25mm}
     \begin{subfigure}{0.95\textwidth}
        \centering
        \includegraphics[width=\textwidth]{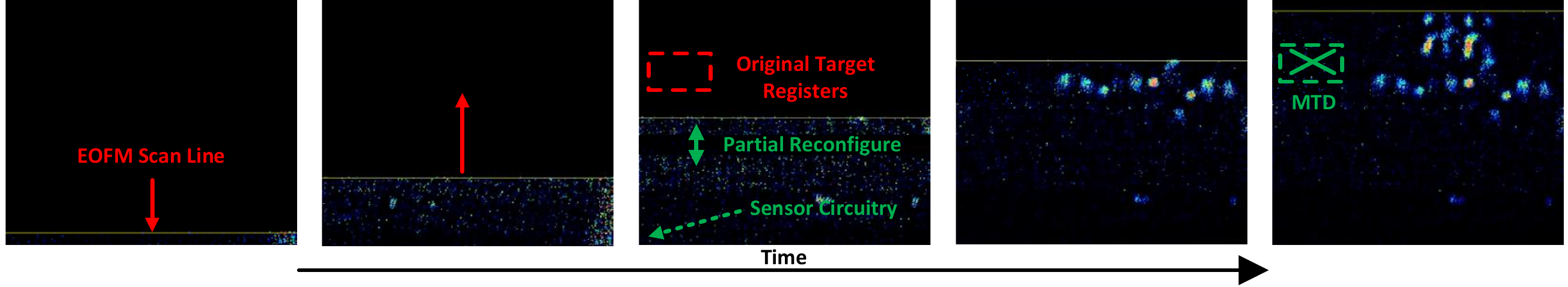}
        \vspace*{-6mm}
        \caption{timeline illustration of EOFM scan to extract targeting register values while \textit{LaserEscape} activated.}
        \label{fig:eofm_def}
     \end{subfigure}
        \caption{EOFM probing attack timeline on images  captured during laser scanning from back-side of the FPGA.}
        \label{fig:eofm_res}
\end{figure*}



\noindent\textbf{Device Under Test.}\label{subsec:DUT}
For our experiments we used Digilent's Genesys 2 development board, equipped with a Kintex-7 FPGA from Xilinx/AMD (part number XC7K325T-2FFG900C). 
We developed a custom-made 3D printed stage to hold the target device and the controller system.

\noindent\textbf{Communication.} For the run-time configuration/communication of the target FPGA, we used a 3.3V modified Arduino Uno R3 communicating over I$^2$C interface with the FPGA target. A compact circuitry on the FPGA part is deployed to handle I$^2$C communication and commands from the user. We choose an I$^2$C bus since the addressable communicated data makes adding post-processing circuits easy for different experiments.

\noindent\textbf{Bitstream Reconfiguration.} For the MTD implantation, once the target is initiated with the mitigation mechanism, we utilize a Raspberry Pi 3 Model B equipped with a Quad Core 1.2GHz Broadcom 64bit CPU for both partial reconfiguration and programming.

\noindent\textbf{Analyzer and Controller Configuration.} 
Illustrated in Fig.~\ref{fig:experiment_set}, we use a dedicated controller system, to send commands, program key registers, adjusting target clock, and handle I/O for experiments via I$^2$C. The measurement system is a dedicated system synced with the \textit{PHEMOS-X} to control the stage, laser configurations and display the digitized data captured from the detection RAC. We use \textit{Semishop Failure Analysis Software 4.0} to control the stage and capture the images and signals.


\noindent\textbf{Optical Probing Configuration.}
For the EOFM and EOP analysis we used 20x objective lens to focus on the target circuitry. The EOFM target frequency in our experiments is set to \textit{1.25MHz}, meaning that we use the controlling system to alternate the global \textit{Rst} of the target every $200 \mu s$. The integration time for the EOFM scan speed is set to 1 ms/px with \textit{1KHz} of bandwidth. For the EOP, the probing resolution is set to 100ps with 10,000 iterations. With the aforementioned configurations, a complete EOFM scan in 1x zoom takes approximately 194 seconds and as for a complete integrated EOP it takes 24 seconds.

\noindent\textbf{Sensor Configuration.}
After implementing the target RTL, we added the sensor in a neighbour SLICE of the target. In our sensor implementation, we used a chain of 8-long IDELAYE2 ($m=8$) for clock delay path. The tuning time interval is set to $t_{sense}=100ms$ and the zero counter interval window in test time is set to $t_{detect}=255$ clock cycles. We chose $m$ and $t_{detect}$ because they were the largest values we could read or write with a single byte over I$^2$C bus. We chose $t_{sense}$ because to achieve a balance of sensitivity versus speed when conducting experiments.


\section{Results}\label{sec:eval}

\subsection{EOFM for Localizing Target Registers}\label{sec:eofm_attack}
For the EOFM attacks on the key registers, we follow the similar attack flow described in~\cite{lohrke2016no}. Specifically, we mount a profiling attack on key registers implemented by flip-flop with clock enable and synchronous reset (\textit{FDRE})~\cite{ultrascale}. We carry out a localization attack accurately by exploiting the \textit{reset} signal of the DUT~\cite{lohrke2016no}. In this attack, the adversary can localize the coarse-grain circuitry of interest by feeding the known values into the chip via \textit{I/O} with a particular frequency. Then, by sweeping in a particular frequency bandwidth with the center frequency of \textit{I/O} activities, she can effectively discover the target registers and the corresponding connecting circuitry near them on the chip. Fig.~\ref{fig:eofm_un} shows the timeline of an EOFM scan on a target area. As shown, the target registers are accurately localized on the selected frequency (i.e., \textit{reset} frequency) once the EOFM scanner line reaches the area. 

Images captured from back-side FPGA in Fig.~\ref{fig:eofm_un} show how the scanning is carried out (from left to right). As the horizontal scan line (across Y axis) moves, details of the internals of the FPGA are shown. The bright spots represent the electronic elements (e.g., registers) that actively operate on particular frequencies within the range of the EOFM scan. The target register areas, highlighted in a red box, are programmed registers that are localized by profiling. Similar to methods in~\cite{lohrke2016no}, the profiling process consists of 1) programming the elements to fluctuate in a certain frequency, 2) activate/deactivate them and 3) analyze the optical activity extracted from EOFM. In our case shown in Fig.~\ref{fig:eofm_un}, 1) the target \textit{FDRE} registers are programmed to reset with EOFM scanning frequency, 2) are set and clear in multiple iterations and 3) are inspected visually after EOFM scan in terms of the optical activity. The bright area within the red target area indicates that 1) the registers are fluctuating on the particular frequency and 2) they are set to one.


\subsection{EOP for Register Values}

\begin{figure}[!h]
   \centering \noindent
   \includegraphics[width=.95\linewidth]{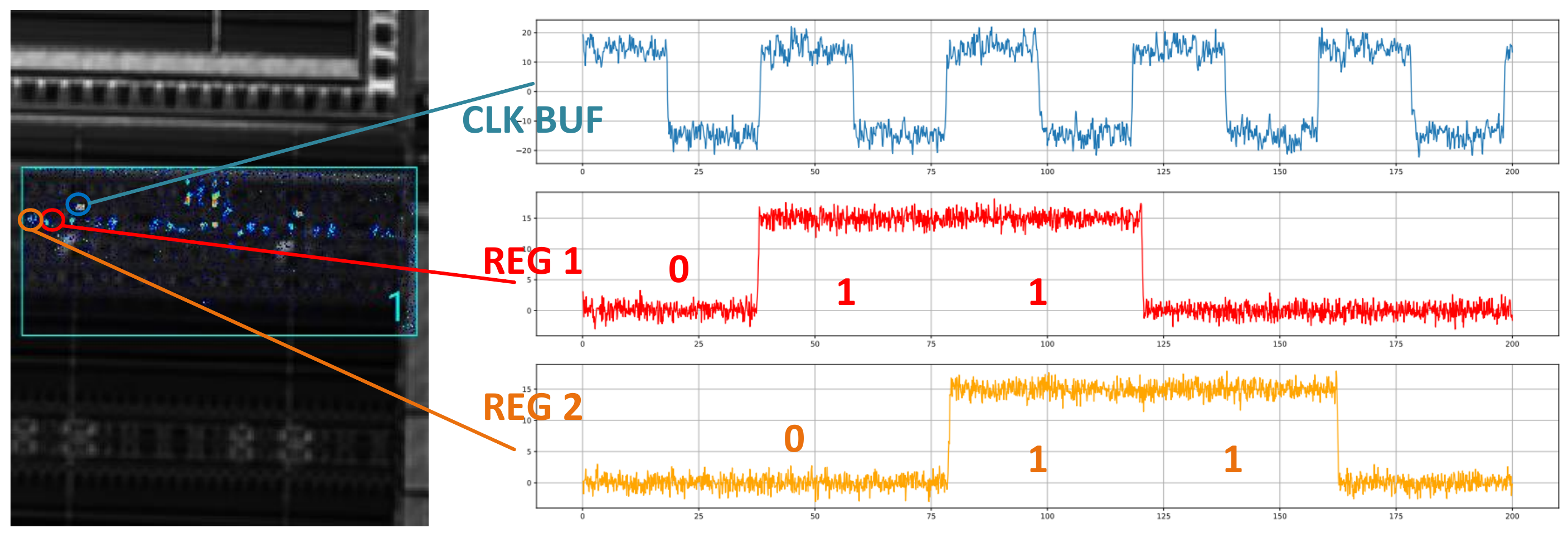}
	\caption{EOFM and corresponding EOP waveforms on target registers. }
   	\label{fig:eop_tar}
 \end{figure}

Upon successful localization of the registers, we can then mount an EOP attack by focusing the HIL beam on each target FF. This process can be executed during the secret operation of the IC. For instance, if the state registers of a block cipher (e.g., AES) are probed, the keys are revealed easily. Fig~\ref{fig:eop_tar} showcases a successful EOP attack on an implementation of shift-register. 
In this particular scenario, we focus the lens onto the registers (which are already localized via EOFM) and carry out the shift operation that is programmed on the FPGA. The probing process is synchronized with the operations on the FPGA via triggers. The probing experiment is performed during a particular interval with picosecond time accuracy to capture fast logical operations which results in voltage alterations in nodes on the chip. After multiple iterations and measurement averaging, the curves illustrated in Fig~\ref{fig:eop_tar} are produced. Sometimes signal processing techniques, such as frequency band filtering and de-noising, are required to extract the clean states of the targets. This is due to the errors in placing and focusing the lens at the target element. As depicted in Fig~\ref{fig:eop_tar}, the shift register operations in \textit{REG 1} and \textit{REG 2} can be followed synchronously with the \textit{CLK-BUF} rising edges. 

\subsection{EOFM for Function Recovery }\label{function_rec}

\begin{figure}[!t]
   \centering \noindent
   \includegraphics[width=0.7\linewidth]{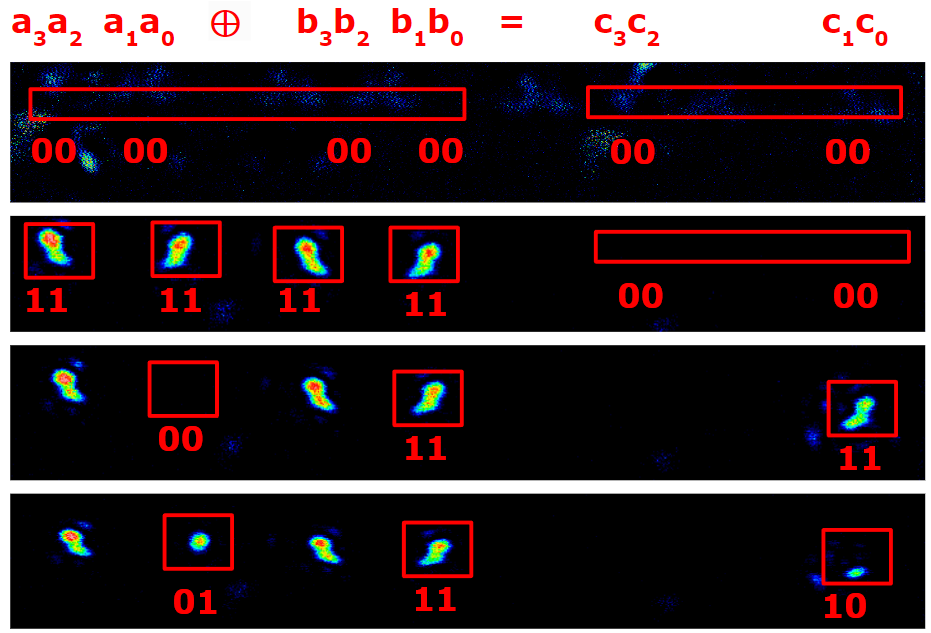}
	\caption{Set of EOFM images for function(XOR) recovery}
   	\label{fig:eofm_xor}
 \end{figure}

As an extension to the key recovery attempt, we also demonstrate a function recovery attack via EOFM. Similar to previous probings, the adversary localizes the region of interest by exploiting the control signals in the frequency domain. 
By conducting a careful profiling procedure, a reverse-engineering approach is possible to discover the functionality of the building blocks. Fig.~\ref{fig:eofm_xor} serves as an example for a series of profiling EOFM patterns of registers in a function. It can be inferred that the correlation of the FFs patterns can be described by an \textit{XOR} function among the 8 input FFs at the left side of the image and the output FFs at the right of the image. As indicated, the programmed functionality for this function is a simple bit-wise \textit{XOR} operation where $c_i = a_i\oplus b_i$. By analysing the EOFM optical response from different input patterns, we can visually confirm the activation patterns for each individual register bits. By a simple observation at the first two rows of Fig.~\ref{fig:eofm_xor}, one can suggest that the response of the neighbouring underlying registers are separated two by two. Particularly as highlighted by red squares, the optical patterns of $a_3a_2$ and $a_1a_0$ are physically separated. This might suggest the physical realizations of FFs in each slice on this particular FPGA architecture. Furthermore by analyzing the last row in Fig.~\ref{fig:eofm_xor}, we can recover the functionality of the operations programmed for this target. Specifically, for $a_1a_0=0b01$, we can observe that that brightness pattern differs from   $0b11$. With the same observation on $c_1c_0=0b10$ we can recover the functionality of the underlying operation in this analysis. 

\subsection{Evaluating Sensor Trigger}
\begin{figure}[!t]
   \centering \noindent
  \includegraphics[width=0.9\linewidth]{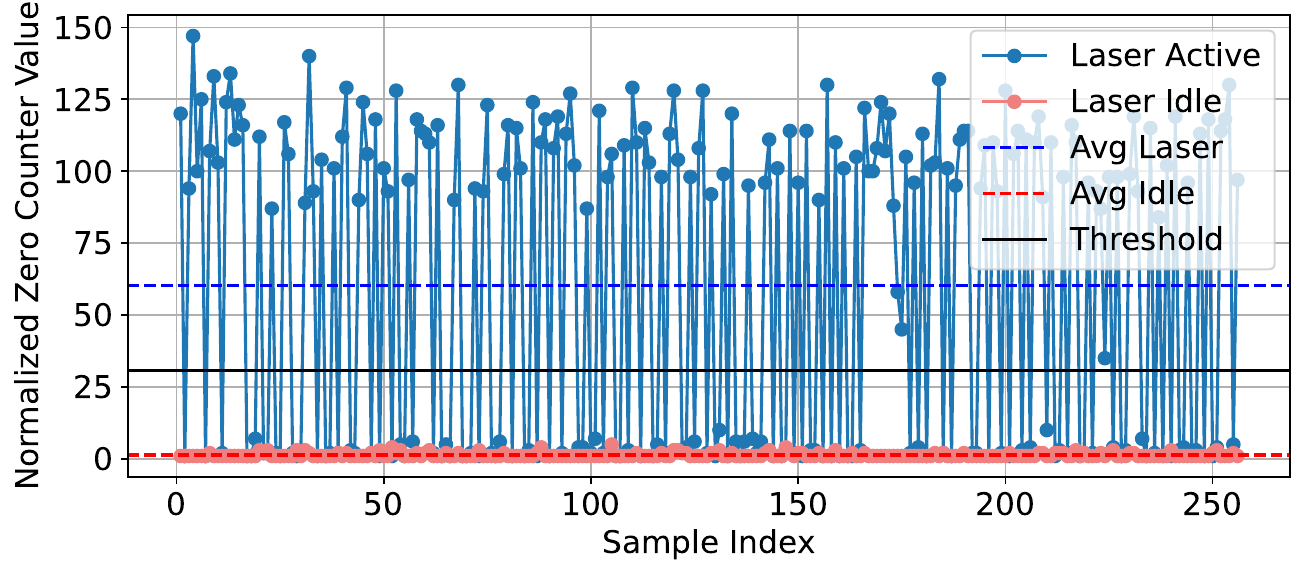}
	\caption{Sensor output against optical probing attempt. }
   	\label{fig:sens_trigger}
 \end{figure}

In order to ensure the sensor operates properly, we empirically investigate the output of the sensor. 
As an example, Fig~\ref{fig:sens_trigger} displays the normalized zero counter output of the sensor in two test cases during a time window of $t_{detect}=255$ clock cycles. 
As it can be observed, while the laser is activated, the average of the zero counter samples is significantly larger compared to the scenario where the laser is idle. Furthermore, Fig~\ref{fig:sens_trigger} shows that in this case, the trigger threshold can be selected by a safe margin. 
Note that compared to the EOFM/EOP integration time, the highlighted time window where the sensor samples are collected is very short. Specifically, for a $f=100MHz$ operation frequency, the samples are collected during $256\times 10^{-8}=2.56 \mu s$, which is considerably faster compared to the dwell time of the laser during probing.
 



\subsection{Defeating Function Discovery Attacks}
\begin{figure}[!t]
   \centering \noindent
   \includegraphics[width=0.8\linewidth]{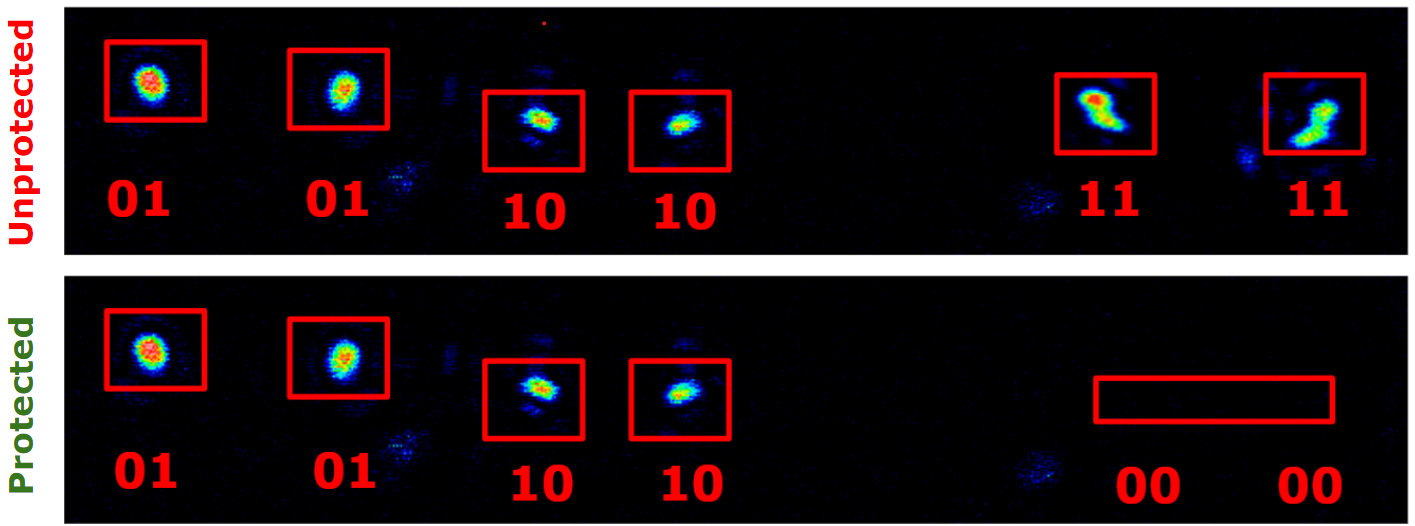}
	\caption{ EOFM images for function(XOR) recovery and polymorphic gate mitigation.}
   	\label{fig:poly_res}
 \end{figure}

For this experiment, we deploy \textit{LaserEscape} sensor on the target FPGA. We mount the same optical probing attack described in Sect.~\ref{function_rec}. In this experiment instead of a simple \textsc{XOR} gates for the target function we deployed the proposed polymorphic \textit{XOR} gates in the target design. For the EOFM, Fig.~\ref{fig:poly_res} displays the input and output registers of the target function. As shown in the upper part of Fig.~\ref{fig:poly_res}, the unprotected target reveals the \textit{XOR} function for $in_1=0b0101$ and $in_2=0b1010$ which results in $out=0b1111$ highlighted on the right side of the figure. When the sensor detects laser emission on the FPGA, the polymorphic gate activates and clears the output of the \textit{XOR} gates. This is highlighted in the lower part of Fig.~\ref{fig:poly_res}.

\subsection{Defeating Register Recovery}
To evaluate our method we organize the similar EOFM attack described in Sect.~\ref{sec:eofm_attack} in presence of \textit{LaserEscape}. We target a set of 8 FDRE registers which are considered to be protected in the design. One can assume these registers hold the master key of a cryptographic algorithm (e.g., AES). Fig.~\ref{fig:eofm_def} illustrated the timeline of the EOFM scan targeting the FDRE registers. Compared to the similar attack depicted in Fig.\ref{fig:eofm_un}, as the scanning laser line approaches the physical locations of the target FDREs, the sensor in  \textit{LaserEscape} is triggered and consequently the MTD via PR is carried out in real-time, leaving no trace of the FDRE targets once the laser arrives at the target destination. Note that the sensor circuitry is placed on a neighbouring slice to the target shown in Fig.~\ref{fig:eofm_def} and is not illustrated in these captures.





 \section{Discussions}\label{sec:dis}


\subsection{Overhead Analysis}
In terms of the overhead introduced by \textit{LaserEscape}, the 1LUTsensor is very efficient in terms of area overhead by employing a single LUT and utilizing 9 IDELAYE2 in the design, which is negligible. 
For the MTD, if 32-bit state registers of an AES block cipher is considered as the target, a total area overhead of 3.5\% (compared to original AES) is estimated by \textit{Vivado} toolkit in the case of choosing 8 Reconfigurable LOCs. The delay overhead caused by the PR process for the same implementation is measured to be $223\mu s$, which maintains a large safe gap against the laser dwell time per pixel for EOP and EOFM attacks.

\subsection{Stability}\label{sec:stability}
\begin{figure}[!h]
    \centering
    \includegraphics[width=0.45\textwidth]{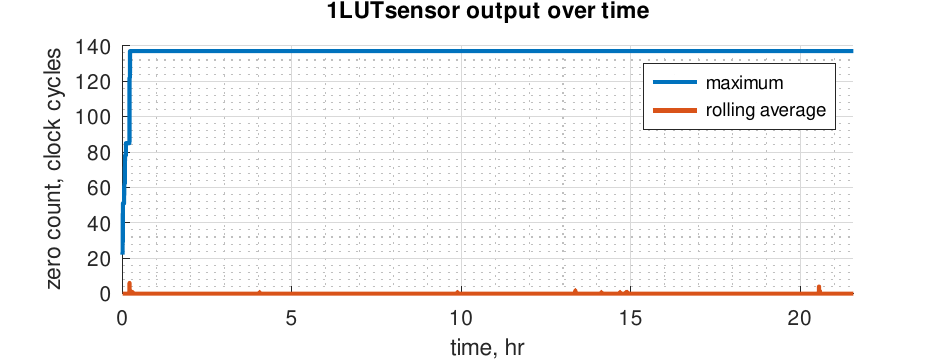}
    \caption{Plot of maximum and rolling average zero counts over 21.5 hours}
    \label{fig:driftgraph}
\end{figure}

As long as the ambient conditions and FPGA load remain the same, our sensor remains stable. We let the sensor run idle for about 21 hours, logging the sensor's readings over  zero counts. Our experiments, shown in Fig.~\ref{fig:driftgraph}, show that the maximum zero count had plateaued after less than 20 minutes, but the rolling average stayed at around zero for the rest of 21.5 hours. This implies that the zero counts follow a normal distribution and the sensor is stable.


\vspace*{-5mm}

\subsection{Further Improvements}
It is possible to statistically investigate the area coverage of the delay sensor with regards to emitted laser power, scan duration, corresponding connections of the sensor circuitry (e.i,  wires), and different tuning of the \textit{IDELAYs} or any similar delay-based sensor against optical attacks. It would also be worthy to study the possible improvement of detection area coverage if multiple sensors are deployed in the target circuit.  


 \section{Conclusion}\label{sec:Conclusion}
In this paper, we presented the first fully digital and FPGA-compatible detection and response countermeasure, called \textit{LaserEscape}, against optical probing attacks.
\textit{LaserEscape} leverages delay-based sensors to detect optical probing attempts and MTD strategies through the PR feature of conventional FPGAs and polymorphic circuits to respond to the attack.
By repurposing the 1LUTsensor, deploying open-source bitstream manipulators, and designing LUT-based polymorphic gates, we built a real-time countermeasure to detect the laser beam and deceive the adversary.
We demonstrated the resiliency of our approach by mounting optical probing attacks against target implementations on a 28 nm FPGA in two scenarios, namely key extraction and function extraction.

\section*{Acknowledgment}
This effort was sponsored by NSF Grants CNS-2150123 and CNS-2150122.


\bibliographystyle{ACM-Reference-Format}

\bibliography{ref}


\begin{thebibliography}{40}


\ifx \showCODEN    \undefined \def \showCODEN     #1{\unskip}     \fi
\ifx \showDOI      \undefined \def \showDOI       #1{#1}\fi
\ifx \showISBNx    \undefined \def \showISBNx     #1{\unskip}     \fi
\ifx \showISBNxiii \undefined \def \showISBNxiii  #1{\unskip}     \fi
\ifx \showISSN     \undefined \def \showISSN      #1{\unskip}     \fi
\ifx \showLCCN     \undefined \def \showLCCN      #1{\unskip}     \fi
\ifx \shownote     \undefined \def \shownote      #1{#1}          \fi
\ifx \showarticletitle \undefined \def \showarticletitle #1{#1}   \fi
\ifx \showURL      \undefined \def \showURL       {\relax}        \fi
\providecommand\bibfield[2]{#2}
\providecommand\bibinfo[2]{#2}
\providecommand\natexlab[1]{#1}
\providecommand\showeprint[2][]{arXiv:#2}

\bibitem[Bernard et~al\mbox{.}(2021)]%
        {bernard2021design}
\bibfield{author}{\bibinfo{person}{Chandler Bernard}, \bibinfo{person}{William Bryant}, \bibinfo{person}{Richard Becker}, {and} \bibinfo{person}{Jia Di}.} \bibinfo{year}{2021}\natexlab{}.
\newblock \showarticletitle{Design of Asynchronous Polymorphic Logic Gates for Hardware Security}. In \bibinfo{booktitle}{\emph{2021 IEEE High Performance Extreme Computing Conference (HPEC)}}. IEEE, \bibinfo{pages}{1--5}.
\newblock


\bibitem[Bow et~al\mbox{.}(2020)]%
        {bow2020side}
\bibfield{author}{\bibinfo{person}{Ivan Bow}, \bibinfo{person}{Nahome Bete}, \bibinfo{person}{Fareena Saqib}, \bibinfo{person}{Wenjie Che}, \bibinfo{person}{Chintan Patel}, \bibinfo{person}{Ryan Robucci}, \bibinfo{person}{Calvin Chan}, {and} \bibinfo{person}{Jim Plusquellic}.} \bibinfo{year}{2020}\natexlab{}.
\newblock \showarticletitle{Side-channel power resistance for encryption algorithms using implementation diversity}.
\newblock \bibinfo{journal}{\emph{Cryptography}} \bibinfo{volume}{4}, \bibinfo{number}{2} (\bibinfo{year}{2020}), \bibinfo{pages}{13}.
\newblock


\bibitem[Cannon et~al\mbox{.}(2023)]%
        {cannon2023protection}
\bibfield{author}{\bibinfo{person}{Andrew Cannon}, \bibinfo{person}{Tasnuva Farheen}, \bibinfo{person}{Sourav Roy}, \bibinfo{person}{Shahin Tajik}, {and} \bibinfo{person}{Domenic Forte}.} \bibinfo{year}{2023}\natexlab{}.
\newblock \showarticletitle{Protection against physical attacks through self-destructive polymorphic latch}. In \bibinfo{booktitle}{\emph{2023 IEEE/ACM International Conference on Computer Aided Design (ICCAD)}}. IEEE, \bibinfo{pages}{1--9}.
\newblock


\bibitem[Chef et~al\mbox{.}(2018)]%
        {chef2018descrambling}
\bibfield{author}{\bibinfo{person}{Samuel Chef}, \bibinfo{person}{Chung~Tah Chua}, \bibinfo{person}{Jing~Yun Tay}, \bibinfo{person}{Yu~Wen Siah}, \bibinfo{person}{Shivam Bhasin}, \bibinfo{person}{J Breier}, {and} \bibinfo{person}{Chee~Lip Gan}.} \bibinfo{year}{2018}\natexlab{}.
\newblock \showarticletitle{Descrambling of embedded {SRAM} using a laser probe}. In \bibinfo{booktitle}{\emph{2018 IEEE International Symposium on the Physical and Failure Analysis of Integrated Circuits (IPFA)}}. IEEE, \bibinfo{pages}{1--6}.
\newblock


\bibitem[Falkinburg(2011)]%
        {Falkinburg_2011}
\bibfield{author}{\bibinfo{person}{Jeffery Falkinburg}.} \bibinfo{year}{2011}\natexlab{}.
\newblock \emph{\bibinfo{title}{Dynamic Polymorphic Reconfiguration to Effectively Cloak a Circuit’s Function}}.
\newblock \bibinfo{thesistype}{Ph.\,D. Dissertation}. \bibinfo{school}{Air Force Institute of Technology}.
\newblock


\bibitem[Farheen et~al\mbox{.}(2022)]%
        {farheen2022twofold}
\bibfield{author}{\bibinfo{person}{Tasnuva Farheen}, \bibinfo{person}{Sourav Roy}, \bibinfo{person}{Shahin Tajik}, {and} \bibinfo{person}{Domenic Forte}.} \bibinfo{year}{2022}\natexlab{}.
\newblock \showarticletitle{A twofold clock and voltage-based detection method for laser logic state imaging attack}.
\newblock \bibinfo{journal}{\emph{IEEE Transactions on Very Large Scale Integration (VLSI) Systems}} \bibinfo{volume}{31}, \bibinfo{number}{1} (\bibinfo{year}{2022}), \bibinfo{pages}{65--78}.
\newblock


\bibitem[Gro{\ss} et~al\mbox{.}(2016)]%
        {gross2016domain}
\bibfield{author}{\bibinfo{person}{Hannes Gro{\ss}}, \bibinfo{person}{Stefan Mangard}, {and} \bibinfo{person}{Thomas Korak}.} \bibinfo{year}{2016}\natexlab{}.
\newblock \showarticletitle{Domain-oriented masking: Compact masked hardware implementations with arbitrary protection order}.
\newblock \bibinfo{journal}{\emph{Cryptology ePrint Archive}} (\bibinfo{year}{2016}).
\newblock


\bibitem[G{\"u}neysu and Moradi(2011)]%
        {guneysu2011generic}
\bibfield{author}{\bibinfo{person}{Tim G{\"u}neysu} {and} \bibinfo{person}{Amir Moradi}.} \bibinfo{year}{2011}\natexlab{}.
\newblock \showarticletitle{Generic side-channel countermeasures for reconfigurable devices}. In \bibinfo{booktitle}{\emph{International workshop on cryptographic hardware and embedded systems}}. Springer, \bibinfo{pages}{33--48}.
\newblock


\bibitem[Hettwer et~al\mbox{.}(2019)]%
        {hettwer2019securing}
\bibfield{author}{\bibinfo{person}{Benjamin Hettwer}, \bibinfo{person}{Johannes Petersen}, \bibinfo{person}{Stefan Gehrer}, \bibinfo{person}{Heike Neumann}, {and} \bibinfo{person}{Tim G{\"u}neysu}.} \bibinfo{year}{2019}\natexlab{}.
\newblock \showarticletitle{Securing cryptographic circuits by exploiting implementation diversity and partial reconfiguration on FPGAs}. In \bibinfo{booktitle}{\emph{2019 Design, Automation \& Test in Europe Conference \& Exhibition (DATE)}}. IEEE, \bibinfo{pages}{260--263}.
\newblock


\bibitem[Heyszl et~al\mbox{.}(2012)]%
        {heyszl2012localized}
\bibfield{author}{\bibinfo{person}{Johann Heyszl}, \bibinfo{person}{Stefan Mangard}, \bibinfo{person}{Benedikt Heinz}, \bibinfo{person}{Frederic Stumpf}, {and} \bibinfo{person}{Georg Sigl}.} \bibinfo{year}{2012}\natexlab{}.
\newblock \showarticletitle{Localized electromagnetic analysis of cryptographic implementations}. In \bibinfo{booktitle}{\emph{Topics in Cryptology--CT-RSA 2012: The Cryptographers’ Track at the RSA Conference 2012, San Francisco, CA, USA, February 27--March 2, 2012. Proceedings}}. Springer, \bibinfo{pages}{231--244}.
\newblock


\bibitem[Jayasinghe et~al\mbox{.}(2023)]%
        {Jayasinghe_Udugama_Parameswaran_2023}
\bibfield{author}{\bibinfo{person}{Darshana Jayasinghe}, \bibinfo{person}{Brian Udugama}, {and} \bibinfo{person}{Sri Parameswaran}.} \bibinfo{year}{2023}\natexlab{}.
\newblock \showarticletitle{1LUTSensor: Detecting FPGA Voltage Fluctuations using LookUp Tables}.
\newblock \bibinfo{journal}{\emph{IACR Transactions on Cryptographic Hardware and Embedded Systems}}  \bibinfo{volume}{2024} (\bibinfo{date}{Dec.} \bibinfo{year}{2023}), \bibinfo{pages}{51–86}.
\newblock
\urldef\tempurl%
\url{https://doi.org/10.46586/tches.v2024.i1.51-86}
\showDOI{\tempurl}


\bibitem[Khan et~al\mbox{.}(2021)]%
        {khan2021moving}
\bibfield{author}{\bibinfo{person}{Nadir Khan}, \bibinfo{person}{Benjamin Hettwer}, {and} \bibinfo{person}{J{\"u}rgen Becker}.} \bibinfo{year}{2021}\natexlab{}.
\newblock \showarticletitle{Moving target and implementation diversity based countermeasures against side-channel attacks}. In \bibinfo{booktitle}{\emph{International Symposium on Applied Reconfigurable Computing}}. Springer, \bibinfo{pages}{188--202}.
\newblock


\bibitem[Koch(2012)]%
        {koch2012partial1}
\bibfield{author}{\bibinfo{person}{Dirk Koch}.} \bibinfo{year}{2012}\natexlab{}.
\newblock \bibinfo{booktitle}{\emph{Partial reconfiguration on FPGAs: architectures, tools and applications}}. Vol.~\bibinfo{volume}{153}.
\newblock \bibinfo{publisher}{Springer Science \& Business Media}.
\newblock


\bibitem[Koch et~al\mbox{.}(2012)]%
        {koch2012partial}
\bibfield{author}{\bibinfo{person}{Dirk Koch}, \bibinfo{person}{Jim Torresen}, \bibinfo{person}{Christian Beckhoff}, \bibinfo{person}{Daniel Ziener}, \bibinfo{person}{Christopher Dennl}, \bibinfo{person}{Volker Breuer}, \bibinfo{person}{J{\"u}rgen Teich}, \bibinfo{person}{Michael Feilen}, {and} \bibinfo{person}{Walter Stechele}.} \bibinfo{year}{2012}\natexlab{}.
\newblock \showarticletitle{Partial reconfiguration on FPGAs in practice—Tools and applications}. In \bibinfo{booktitle}{\emph{ARCS 2012}}. IEEE, \bibinfo{pages}{1--12}.
\newblock


\bibitem[Krachenfels et~al\mbox{.}(2021a)]%
        {krachenfels2021real}
\bibfield{author}{\bibinfo{person}{Thilo Krachenfels}, \bibinfo{person}{Fatemeh Ganji}, \bibinfo{person}{Amir Moradi}, \bibinfo{person}{Shahin Tajik}, {and} \bibinfo{person}{Jean-Pierre Seifert}.} \bibinfo{year}{2021}\natexlab{a}.
\newblock \showarticletitle{Real-world snapshots vs. theory: Questioning the t-probing security model}. In \bibinfo{booktitle}{\emph{2021 IEEE symposium on security and privacy (SP)}}. IEEE, \bibinfo{pages}{1955--1971}.
\newblock


\bibitem[Krachenfels et~al\mbox{.}(2021b)]%
        {krachenfels2021automatic}
\bibfield{author}{\bibinfo{person}{Thilo Krachenfels}, \bibinfo{person}{Tuba Kiyan}, \bibinfo{person}{Shahin Tajik}, {and} \bibinfo{person}{Jean-Pierre Seifert}.} \bibinfo{year}{2021}\natexlab{b}.
\newblock \showarticletitle{Automatic Extraction of Secrets from the Transistor Jungle using $\{$Laser-Assisted$\}$$\{$Side-Channel$\}$ Attacks}. In \bibinfo{booktitle}{\emph{30th USENIX security symposium (USENIX security 21)}}. \bibinfo{pages}{627--644}.
\newblock


\bibitem[Lohrke et~al\mbox{.}(2016)]%
        {lohrke2016no}
\bibfield{author}{\bibinfo{person}{Heiko Lohrke}, \bibinfo{person}{Shahin Tajik}, \bibinfo{person}{Christian Boit}, {and} \bibinfo{person}{Jean-Pierre Seifert}.} \bibinfo{year}{2016}\natexlab{}.
\newblock \showarticletitle{No place to hide: Contactless probing of secret data on FPGAs}. In \bibinfo{booktitle}{\emph{Cryptographic Hardware and Embedded Systems--CHES 2016: 18th International Conference, Santa Barbara, CA, USA, August 17-19, 2016, Proceedings 18}}. Springer, \bibinfo{pages}{147--167}.
\newblock


\bibitem[Manev et~al\mbox{.}(2022)]%
        {manev2022byteman}
\bibfield{author}{\bibinfo{person}{Kristiyan Manev}, \bibinfo{person}{Joseph Powell}, \bibinfo{person}{Kaspar Matas}, {and} \bibinfo{person}{Dirk Koch}.} \bibinfo{year}{2022}\natexlab{}.
\newblock \showarticletitle{byteman: A Bitstream Manipulation Framework}. In \bibinfo{booktitle}{\emph{2022 International Conference on Field-Programmable Technology (ICFPT)}}. IEEE, \bibinfo{pages}{1--9}.
\newblock


\bibitem[Mehta et~al\mbox{.}(2024)]%
        {mehta20241}
\bibfield{author}{\bibinfo{person}{Dev~M Mehta}, \bibinfo{person}{Mohammad Hashemi}, \bibinfo{person}{Domenic Forte}, \bibinfo{person}{Shahin Tajik}, {and} \bibinfo{person}{Fatemeh Ganji}.} \bibinfo{year}{2024}\natexlab{}.
\newblock \showarticletitle{1/0 Shades of UC: Photonic Side-Channel Analysis of Universal Circuits}.
\newblock \bibinfo{journal}{\emph{Cryptology ePrint Archive}} (\bibinfo{year}{2024}).
\newblock


\bibitem[Mentens et~al\mbox{.}(2008)]%
        {mentens2008power}
\bibfield{author}{\bibinfo{person}{Nele Mentens}, \bibinfo{person}{Benedikt Gierlichs}, {and} \bibinfo{person}{Ingrid Verbauwhede}.} \bibinfo{year}{2008}\natexlab{}.
\newblock \showarticletitle{Power and fault analysis resistance in hardware through dynamic reconfiguration}. In \bibinfo{booktitle}{\emph{International Workshop on Cryptographic Hardware and Embedded Systems}}. Springer, \bibinfo{pages}{346--362}.
\newblock


\bibitem[Monfared et~al\mbox{.}(2024)]%
        {monfared2024randohm}
\bibfield{author}{\bibinfo{person}{Saleh~Khalaj Monfared}, \bibinfo{person}{Domenic Forte}, {and} \bibinfo{person}{Shahin Tajik}.} \bibinfo{year}{2024}\natexlab{}.
\newblock \showarticletitle{RandOhm: Mitigating Impedance Side-channel Attacks using Randomized Circuit Configurations}.
\newblock \bibinfo{journal}{\emph{arXiv preprint arXiv:2401.08925}} (\bibinfo{year}{2024}).
\newblock


\bibitem[Monfared et~al\mbox{.}(2020)]%
        {monfared2020bsrng}
\bibfield{author}{\bibinfo{person}{Saleh~Khalaj Monfared}, \bibinfo{person}{Omid Hajihassani}, \bibinfo{person}{Mohammad~Sina Kiarostami}, \bibinfo{person}{Soroush~Meghdadi Zanjani}, \bibinfo{person}{Dara Rahmati}, {and} \bibinfo{person}{Saeid Gorgin}.} \bibinfo{year}{2020}\natexlab{}.
\newblock \showarticletitle{BSRNG: a high throughput parallel bitsliced approach for random number generators}. In \bibinfo{booktitle}{\emph{Workshop Proceedings of the 49th International Conference on Parallel Processing}}. \bibinfo{pages}{1--10}.
\newblock


\bibitem[Moradi and Mischke(2013)]%
        {moradi2013comprehensive}
\bibfield{author}{\bibinfo{person}{Amir Moradi} {and} \bibinfo{person}{Oliver Mischke}.} \bibinfo{year}{2013}\natexlab{}.
\newblock \showarticletitle{Comprehensive evaluation of AES dual ciphers as a side-channel countermeasure}. In \bibinfo{booktitle}{\emph{Information and Communications Security: 15th International Conference, ICICS 2013, Beijing, China, November 20-22, 2013. Proceedings 15}}. Springer, \bibinfo{pages}{245--258}.
\newblock


\bibitem[Nikova et~al\mbox{.}(2006)]%
        {nikova2006threshold}
\bibfield{author}{\bibinfo{person}{Svetla Nikova}, \bibinfo{person}{Christian Rechberger}, {and} \bibinfo{person}{Vincent Rijmen}.} \bibinfo{year}{2006}\natexlab{}.
\newblock \showarticletitle{Threshold implementations against side-channel attacks and glitches}. In \bibinfo{booktitle}{\emph{International conference on information and communications security}}. Springer, \bibinfo{pages}{529--545}.
\newblock


\bibitem[Parvin et~al\mbox{.}(2023)]%
        {parvin2023hidden}
\bibfield{author}{\bibinfo{person}{Sajjad Parvin}, \bibinfo{person}{Chandan~Kumar Jha}, \bibinfo{person}{Sallar Ahmadi-Pours}, \bibinfo{person}{Frank~Sill Torres}, {and} \bibinfo{person}{Rolf Drechsler}.} \bibinfo{year}{2023}\natexlab{}.
\newblock \showarticletitle{Hidden in Plain Sight: A Detailed Investigation of Selectively Increasing Local Density to Camouflage and Robustify Against Optical Probing Attacks}. In \bibinfo{booktitle}{\emph{2023 IEEE International Test Conference India (ITC India)}}. IEEE, \bibinfo{pages}{1--6}.
\newblock


\bibitem[Rahman et~al\mbox{.}(2021)]%
        {rahman2021concealing}
\bibfield{author}{\bibinfo{person}{M~Tanjidur Rahman}, \bibinfo{person}{Nusrat~Farzana Dipu}, \bibinfo{person}{Dhwani Mehta}, \bibinfo{person}{Shahin Tajik}, \bibinfo{person}{Mark Tehranipoor}, {and} \bibinfo{person}{Navid Asadizanjani}.} \bibinfo{year}{2021}\natexlab{}.
\newblock \showarticletitle{Concealing-gate: Optical contactless probing resilient design}.
\newblock \bibinfo{journal}{\emph{ACM Journal on Emerging Technologies in Computing Systems (JETC)}} \bibinfo{volume}{17}, \bibinfo{number}{3} (\bibinfo{year}{2021}), \bibinfo{pages}{1--25}.
\newblock


\bibitem[Schellenberg et~al\mbox{.}(2021)]%
        {schellenberg2021inside}
\bibfield{author}{\bibinfo{person}{Falk Schellenberg}, \bibinfo{person}{Dennis~RE Gnad}, \bibinfo{person}{Amir Moradi}, {and} \bibinfo{person}{Mehdi~B Tahoori}.} \bibinfo{year}{2021}\natexlab{}.
\newblock \showarticletitle{An inside job: Remote power analysis attacks on FPGAs}.
\newblock \bibinfo{journal}{\emph{IEEE Design \& Test}} \bibinfo{volume}{38}, \bibinfo{number}{3} (\bibinfo{year}{2021}), \bibinfo{pages}{58--66}.
\newblock


\bibitem[Shen et~al\mbox{.}(2018)]%
        {shen2018nanopyramid}
\bibfield{author}{\bibinfo{person}{Haoting Shen}, \bibinfo{person}{Navid Asadizanjani}, \bibinfo{person}{Mark Tehranipoor}, {and} \bibinfo{person}{Domenic Forte}.} \bibinfo{year}{2018}\natexlab{}.
\newblock \showarticletitle{Nanopyramid: An optical scrambler against backside probing attacks}. In \bibinfo{booktitle}{\emph{ISTFA 2018: Proceedings from the 44th International Symposium for Testing and Failure Analysis}}. ASM International, \bibinfo{pages}{280}.
\newblock


\bibitem[Srivastava and Ghosh(2019)]%
        {srivastava2019efficient}
\bibfield{author}{\bibinfo{person}{Ankush Srivastava} {and} \bibinfo{person}{Prokash Ghosh}.} \bibinfo{year}{2019}\natexlab{}.
\newblock \showarticletitle{An efficient memory zeroization technique under side-channel attacks}. In \bibinfo{booktitle}{\emph{2019 32nd International Conference on VLSI Design and 2019 18th International Conference on Embedded Systems (VLSID)}}. IEEE, \bibinfo{pages}{76--81}.
\newblock


\bibitem[Stoica et~al\mbox{.}(2001)]%
        {stoica2001polymorphic}
\bibfield{author}{\bibinfo{person}{Adrian Stoica}, \bibinfo{person}{Ricardo Zebulum}, {and} \bibinfo{person}{Didier Keymeulen}.} \bibinfo{year}{2001}\natexlab{}.
\newblock \showarticletitle{Polymorphic electronics}. In \bibinfo{booktitle}{\emph{Evolvable Systems: From Biology to Hardware: 4th International Conference, ICES 2001 Tokyo, Japan, October 3--5, 2001 Proceedings 4}}. Springer, \bibinfo{pages}{291--302}.
\newblock


\bibitem[Tada et~al\mbox{.}(2021)]%
        {tada2021design}
\bibfield{author}{\bibinfo{person}{Sho Tada}, \bibinfo{person}{Yuki Yamashita}, \bibinfo{person}{Kohei Matsuda}, \bibinfo{person}{Makoto Nagata}, \bibinfo{person}{Kazuo Sakiyama}, {and} \bibinfo{person}{Noriyuki Miura}.} \bibinfo{year}{2021}\natexlab{}.
\newblock \showarticletitle{Design and concept proof of an inductive impulse self-destructor in sense-and-react countermeasure against physical attacks}.
\newblock \bibinfo{journal}{\emph{Japanese Journal of Applied Physics}} \bibinfo{volume}{60}, \bibinfo{number}{SB} (\bibinfo{year}{2021}), \bibinfo{pages}{SBBL01}.
\newblock


\bibitem[Tajik et~al\mbox{.}(2017a)]%
        {tajik2017pufmon}
\bibfield{author}{\bibinfo{person}{Shahin Tajik}, \bibinfo{person}{Julian Fietkau}, \bibinfo{person}{Heiko Lohrke}, \bibinfo{person}{Jean-Pierre Seifert}, {and} \bibinfo{person}{Christian Boit}.} \bibinfo{year}{2017}\natexlab{a}.
\newblock \showarticletitle{Pufmon: Security monitoring of fpgas using physically unclonable functions}. In \bibinfo{booktitle}{\emph{2017 IEEE 23rd International symposium on on-line testing and robust system design (IOLTS)}}. IEEE, \bibinfo{pages}{186--191}.
\newblock


\bibitem[Tajik et~al\mbox{.}(2017b)]%
        {tajik2017power}
\bibfield{author}{\bibinfo{person}{Shahin Tajik}, \bibinfo{person}{Heiko Lohrke}, \bibinfo{person}{Jean-Pierre Seifert}, {and} \bibinfo{person}{Christian Boit}.} \bibinfo{year}{2017}\natexlab{b}.
\newblock \showarticletitle{On the power of optical contactless probing: Attacking bitstream encryption of FPGAs}. In \bibinfo{booktitle}{\emph{Proceedings of the 2017 ACM SIGSAC Conference on Computer and Communications Security}}. \bibinfo{pages}{1661--1674}.
\newblock


\bibitem[Tsoi et~al\mbox{.}(2003)]%
        {tsoi2003compact}
\bibfield{author}{\bibinfo{person}{Kuen~Hung Tsoi}, \bibinfo{person}{Ka~Hei Leung}, {and} \bibinfo{person}{Philip Heng~Wai Leong}.} \bibinfo{year}{2003}\natexlab{}.
\newblock \showarticletitle{Compact FPGA-based true and pseudo random number generators}. In \bibinfo{booktitle}{\emph{11th Annual IEEE Symposium on Field-Programmable Custom Computing Machines, 2003. FCCM 2003.}} IEEE, \bibinfo{pages}{51--61}.
\newblock


\bibitem[{Xilinx}(2023)]%
        {xilinx_idelaye2_2023}
\bibfield{author}{\bibinfo{person}{{Xilinx}}.} \bibinfo{year}{2023}\natexlab{}.
\newblock \bibinfo{title}{{IDELAYE2}}.
\newblock
\newblock
\urldef\tempurl%
\url{https://docs.amd.com/r/en-US/ug953-vivado-7series-libraries/IDELAYE2}
\showURL{%
\tempurl}
\newblock
\shownote{publisher: AMD, Inc.}.


\bibitem[Xilinx(2023a)]%
        {ultrascale}
\bibfield{author}{\bibinfo{person}{Xilinx}.} \bibinfo{year}{2023}\natexlab{a}.
\newblock \bibinfo{title}{UltraScale Architecture Libraries Guide (UG974)}.
\newblock \bibinfo{howpublished}{\url{https://docs.amd.com/r/en-US/ug974-vivado-ultrascale-libraries/}}.
\newblock


\bibitem[Xilinx(2023b)]%
        {consXilinx}
\bibfield{author}{\bibinfo{person}{Xilinx}.} \bibinfo{year}{2023}\natexlab{b}.
\newblock \bibinfo{title}{Xilinx Constraints Guide}.
\newblock
\newblock
\urldef\tempurl%
\url{https://www.xilinx.com/xilinx-14/cgd.pdf}
\showURL{%
\tempurl}


\bibitem[Xilinx(2023c)]%
        {xilixreconfig}
\bibfield{author}{\bibinfo{person}{Xilinx}.} \bibinfo{year}{2023}\natexlab{c}.
\newblock \bibinfo{title}{Xilinx Introduction to Dynamic Function eXchange}.
\newblock \bibinfo{howpublished}{\url{https://docs.xilinx.com/r/en-US/ug909-vivado-partial-reconfiguration/Introduction-to-Dynamic-Function-eXchange}}.
\newblock


\bibitem[Zhang et~al\mbox{.}(2023)]%
        {zhang2023laser}
\bibfield{author}{\bibinfo{person}{Hui Zhang}, \bibinfo{person}{Longyang Lin}, \bibinfo{person}{Qiang Fang}, {and} \bibinfo{person}{Massimo Alioto}.} \bibinfo{year}{2023}\natexlab{}.
\newblock \showarticletitle{Laser voltage probing attack detection with 100\% area/time coverage at above/below the bandgap wavelength and fully-automated design}.
\newblock \bibinfo{journal}{\emph{IEEE Journal of Solid-State Circuits}} (\bibinfo{year}{2023}).
\newblock


\bibitem[Zhao and Suh(2018)]%
        {zhao2018fpga}
\bibfield{author}{\bibinfo{person}{Mark Zhao} {and} \bibinfo{person}{G~Edward Suh}.} \bibinfo{year}{2018}\natexlab{}.
\newblock \showarticletitle{FPGA-based remote power side-channel attacks}. In \bibinfo{booktitle}{\emph{2018 IEEE Symposium on Security and Privacy (SP)}}. IEEE, \bibinfo{pages}{229--244}.
\newblock


\end{thebibliography}

\end{document}